\documentclass{amsart}

\usepackage{amsbsy}
\usepackage{graphicx}
\graphicspath{{converted_graphics/}}

\begin{document}

\begin{center}
\LARGE \textbf{Maximum entropy Edgeworth estimates of the number of integer points in polytopes}
\end{center}

\begin{center}
\Large Alexander Barvinok and J.A.Hartigan
\end{center}

\begin{center}
3 August 2010
\end{center}

Abstract: \textit{The number of points $x=(x_1 ,x_2 ,...x_n)$ that lie in an 
integer cube $C$ in $R^n$ and satisfy the constraints $\sum\nolimits_j h_{ij}(x_j )=s_i ,1\le i\le d$ is approximated 
by an Edgeworth-corrected Gaussian formula based on the maximum 
entropy density $p$ on $x \in C$, that satisfies $E\sum\nolimits_j h_{ij}(x_j )=s_i ,1\le i\le d$. Under $p$, the variables $X_1 ,X_2 ,...X_n $ are independent with 
densities of exponential form. Letting $S_i$ denote the random variable  $\sum\nolimits_j h_{ij}(X_j )$, conditional on $S=s, X$ is uniformly distributed over the integers in $C$ that satisfy $S=s$. 
The number of points in $C$ satisfying $S=s$ is $p \{S=s\}\exp (I(p))$ where $I(p)$ is the entropy of the density $p$. 
We estimate $p \{S=s\}$ by $p_Z(s)$, the density at $s$ of the multivariate Gaussian $Z$ with the same first two moments as 
$S$; and when $d$ is large we use in addition an Edgeworth factor that requires the first four 
moments of $S$ under $p$. The asymptotic validity of the Edgeworth-corrected estimate is proved and 
demonstrated for counting contingency tables with given row and column sums as the number of rows and columns 
approaches infinity, and demonstrated for counting the number of graphs with a given degree sequence, 
as the number of vertices approaches infinity.}
\newline\newline


\newline\newline
\noindent
Department of Mathematics, University of Michigan, Ann Arbor, MI 48109-1043,

\noindent USA E-mail address: barvinok@umich.edu
\newline\newline
\noindent Department of Statistics, Yale University, New Haven, CT 06520-8290

\noindent
The research of the first author was partially supported by NSF Grants DMS 
0400617 , DMS 0856640, and a United States - Israel BSF grant 2006377.

\newpage 
\noindent \textbf{1  Maximum entropy estimation of the number of integer points}
\newline\newline
\noindent Let $x=(x_1 ,x_2 ,...x_n )$ be a vector in $R^n$. For arbitrary $R^1 \rightarrow R^1$ functions $h_{ij}$ define $S_i =\sum\nolimits_j h_{ij}(x_j ),1\le i\le d$. Let $Q$ be counting measure 
on a cube $C$ of integers in $R^n$. Consider the surface $S=s$ in $R^n$ 
consisting of points $x$ that satisfy the sums $S_i =\sum\nolimits_j h_{ij}  
(x_j )=s_i ,1\le i\le d$. The volume of the surface $Q \{S=s\}>0$ is the number of points 
that lie in $C$ and in the surface $\{S=s\}$.  For $P_U$ the uniform distribution on the cube $C$, $Q\{S=s\}=P_U\{S=s\}Q\{C\}$.
\newline\newline
 Let $X=(X_1 ,X_2 ,...X_n )$ be $n$ 
random variables uniformly distributed over the cube. Since the random 
variables are independent, the central limit theorem will apply to the sums 
$S_i =\sum\nolimits_j h_{ij}  (X_j )$ under suitable conditions on the $h$. Thus we might approximate the 
probability $P_U \{S=s\}$ by 
$p_Z(s)$, the density at $s$ of a multivariate Gaussian $Z$ with the same first and second moments as 
$S$.
We expect this approximation to work well when the mean of $S$ is close to the 
selected values $s$, but not so well in the tails of the distribution.
Therefore we propose {maximum entropy }Gaussian estimation of the volume using an 
approximating Gaussian with mean value $s$. This procedure is  called \textit
{exponential tilting}; see, for example, [KT03].
\newline

\noindent The entropy of a discrete random variable $X$ having density $p$ (with respect to counting measure) is: 
\begin{equation} I(p)=-E\{\log p(X)\}. \end{equation}

\noindent We find the maximum entropy distribution $P$ described in [J57] , 
with density $p $ on a cube $C$ of integers in $R^n$ satisfying $ES=s$. If there is a density
of exponential form
\begin{equation}
P\{X=x\}=p(x)=\exp \{\sum\nolimits_{ij} \lambda _i  h_{ij} (X_j )+\lambda _0 \}
\end{equation}
 where the $\lambda_i $ are chosen to satisfy the expectations $ES=s$, 
and to ensure that $\sum_{x \in C} p(x) =1$, then this density may be shown to be the unique maximum entropy density subject to the constraints $ES=s$.
\newline\newline Under $P$, the variables $X_1 ,X_2 ,...X_n $ are independent with densities 
\begin{equation}
 p_j (x_j)=\exp \{ \sum\nolimits_i {\lambda _i } h_{ij} (x_j )+\upsilon _j \}.
\end{equation}
 And, conditional on $S=s$, 
 $X$ is uniformly distributed over the integers $x$ in $C$ that satisfy $S=s$, with
\begin{equation}
p(x)=\exp \{ \sum_i \lambda_i s_i+\lambda_0 \}=\exp\{-I(p)\},
\end{equation}
since
\begin{equation}
I(p)=\sum_x[-p(x) \log p(x)]= -E\{\sum\nolimits_{ij} \lambda _i  h_{ij} (X_j )+\lambda _0 \}=-\sum_i \lambda_i s_i-\lambda_0 \
\end{equation}
 \noindent
Thus, for any $x$ that satisfies $S=s$, 
\begin{equation}
Q\{S=s\}=P\{S=s\}/p(x)=P\{S=s\}\exp \{I(p)\}.
\end{equation}
The entropy term in this formula was suggested in some special cases in [B09].
\newline\newline
We again estimate $P\{S=s\}$ by $p_Z(s)$, the density at $s$ of a multivariate 
Gaussian $Z$ with the mean and covariance of $S$. The advantage in using the maximum entropy $P$
is that the mean of the Gaussian is $s$, so that "debiased" estimation takes place at the mean.
\newline\newline\noindent
If the $h$ functions are just multiples, say $h_{ij} (X_j )=A_{ij} X_j $, then 
the maximum entropy density $p $ consists of independent  exponential form 
densities 
\begin{equation}
p_j (x)=\exp \{\theta _j x-c(\theta _j )\} 
\end{equation}
on the $X_j $ with 
canonical parameters $\theta _j =\sum\nolimits_i {\lambda _i } A_{ij} $ and 
expectations $c'(\theta _j )$.  The parameters $\lambda _i $ 
are chosen so that$\sum\nolimits_j A_{ij} c'(\theta _j ) =s_i $.
\newline\newline\noindent
Because $p$ is maximum entropy, the $\theta_j $ may also be characterized [Ba09]
 as the unique maxima of the $p$-entropy 
$\sum\nolimits_j [ c(\theta _j )-\theta _j c'(\theta _j )]$ for a given $\theta$ subject 
to $\sum\nolimits_j {A_{ij} c'(\theta _j )} =s_i $. And then
\begin{equation}
Q\{S=s\}=P\{S=s\}/\prod\limits_j \exp \{\theta _j c'(\theta _j )-c(\theta _j)\} =P\{S=s\}\exp\{I(p)\}.
\end{equation}
\newline\noindent
So far, we have followed the approach in [BH10a] of maximum entropy gaussian approximation.
However, when the number $d$ of sums $S_i$ approaches infinity, and the 
variances of the sums are $O(d)$, the relative error in Gaussian approximation to the true density
for the $i^{th}$ sum will be typically $P\{S_i =s_i \}/p_{Z_i}(s_i)-1=O(1/d)$ and the error in approximating the true density for $d$ sums will be about 
$(1+O(1/d))^d-1=O(1).$ In order to get an accurate approximation we need to consider the Edgeworth corrections to the Gaussian approximation, which use the third and fourth 
cumulants of the $S$ distribution.

In [MW90], McKay and Wormald produced an asymptotic formula for the number 
of near regular graphs on $n$ vertices with $k$ edges, where $k$ is proportional 
to $n$. They derive the formula by a 
saddlepoint approximation to Cauchy's integral for determining a coefficient 
in a generating function. Their generating function turns out to be the characteristic function 
of the sums $S$ appropriate for this problem. The maximum 
entropy Edgeworth approximation generalises their formula to graphs with widely varying degree 
sequences in [BH10b]. The maximum entropy method can also be used to estimate the number of graphs 
with given degree sequences and with additional edge specifications such as 
specified cliques or colorings of the graph.

In [CM05], [GMW06], [CM07], [CGM08], [MG] ,Canfield, Greenhill, McKay, Wormald, and Wang
 extended the Cauchy integral approach to asymptotic 
enumeration of two way contingency tables of integers in which the marginal 
sums are known, with the row sums nearly equal and the column sums nearly 
equal. The integers may be non-negative, or constrained to be 0-1. The  
maximum entropy Edgeworth approximation, (see also  [BH09]), generalises their formulae to the case of varying marginal sums. 
The formulae require the first four moments of certain sums of 
independent random variables. The maximum entropy table entries are independent geometric variables 
when the integers in the tables are non-negative, and independent Bernoulli variables 
when the integers are 0-1. 

The advance in the maximum entropy Edgeworth approximation is that it
provides a unified method for the problems mentioned above, and for generalisations of them,  using a standard statistical approximation,( see for example [K06]), based on the first four moments of sums of independent variables determined by the maximum entropy distributions.
\newline\newline\noindent
Diaconis and Efron [DE85] study the distribution of a chi-square statistic 
for the uniform distribution over contingency tables with fixed margins. The 
number of rows and columns are fixed, but the total count approaches 
infinity. If instead the table entries are bounded, but the numbers of rows 
and columns approach infinity, we expect that a maximum entropy approach 
should yield a valid asymptotic estimate of the distribution. Here the maximum entropy table entries
are integer Gaussians: Gaussian variables, with arbitrary means and 
variances, constrained to be integers.
 
\newpage
\noindent
\textbf{2 The Edgeworth approximation for integer random variables of 
increasing dimensionality.}
\newline\newline\noindent
Let $X_d $ be a sequence of $d-$dimensional integer random variables having 
mean $0$. Suppose that the determinant of the lattice generated by values of 
$X_d $ having positive probability is $\Delta _d $. We wish to
 estimate the probability $P\{X_d =0\}$ using the first four 
moments of $X_d $.
\newline\newline\noindent
Define $Q_a (t)=1 \mbox{ if } \max_i|t_i| \le a, Q_a (t)=0 \mbox { if } \max |t_i| > a.$
\newline\newline
\noindent We use the $d$-dimensioned characteristic function $\phi _d (t)=E(it'X_d 
)$, with $t$ a column vector in $R^d$, and $t'$ the corresponding row vector:
\begin{equation}
P\{X_d =0\}=(2\pi )^{-d}\int {Q_\pi } \phi _d (t).
\end{equation}
The cumulant term $K_d^r (t)$ is the polynomial term of degree $r$ in the 
expansion $\log \phi _d (t)=\sum\nolimits_{r=1}^\infty {\textstyle{i^r} 
\over r!} K_d^r (t)$. Specifically,
\begin{equation}
K_d^2(t)=E(t'X_d )^2,\quad K_d^3 (t)=E(t'X_d)^3,\quad K_d^4 (t)=E(t'X_d )^4-3(K_d^2 (t))^2.
\end{equation}
The variance-covariance matrix $V_d $, is determined by the second cumulant:
\begin{equation} 
\sum\nolimits_{ij} t_i  t_j V_d (i,j)=K_d^2 (t).
\end{equation}

 \noindent Define $\kappa _d^3 =E_d \{K_d^3 (t)^2\}, \kappa _d^4 =E_d K_d^4 (t)$ where the expectation $E_d $ is 
with respect to $t\sim N(0,V_d^{-1} )$, a Gaussian variable with mean $0$ and variance-covariance $V_d^{-1}$. The Edgeworth approximation to $P \{X_d =0\}$ is 
\begin{equation}
\hat {P}\{X_d =0\}=\Delta _d (2\pi )^{-d/2}\vert V_d \vert ^{-1/2}\exp 
(-\kappa _d^3 /72+\kappa _d^4 /24).
\end{equation}
\noindent
The approximation consists of the density at zero of a 
Gaussian with variance-covariance $V_d $, multiplied by an Edgeworth term 
correcting for the departure from Gaussianity.
\newline\newline\noindent
We will use the order of magnitude notation
\begin{eqnarray}
 f(d)=o(g(d))&:&  f(d)/g(d) \to 0 \mbox{ as } d \to \infty,\\
 f(d)=O(g(d)) &:& \limsup_d \vert \frac{f(d)}{g(d)} \vert < \infty.
 \end{eqnarray}
\setcounter{equation}{0}
\renewcommand{\theequation}{\Roman{equation}}
\newline\noindent \textbf{Theorem 1}
Let $E_d$ denote expectation with respect to $t\sim N(0,V_d^{-1} ).$
Suppose that for some $M, \quad \varepsilon =M\sqrt {\log d/d} $,
\begin{eqnarray}
\kappa_d^3=O(1),\quad \kappa_d^4=O(1),\\
E_d\left\{Q_\varepsilon \exp[{1 \over 12}K_d^4]\right\}=O(1),\\
  Q_\varepsilon[ \log \phi _d (t)-\sum\nolimits_{r=2}^4 K_d^r(t){i^r \over {r!}}] &=& o(1) , \\
E_d \left\{Q_\varepsilon \exp [-{1 \over 6}iK_d^3 (t)+{1 \over 72}\kappa _d^3+{1 \over 24}K_d^4 
(t)-{1 \over 24}\kappa _d^4]\right\}&\to&1 \mbox{ as } d \to \infty,\\
\int_ {Q_\pi -Q_\varepsilon } \vert \phi _d (t)\vert /\int_ 
{Q_\varepsilon} |\phi _d (t)|&=&o(1).
\end{eqnarray}
\renewcommand{\theequation}{\arabic{equation}}
\setcounter{equation}{12}
\begin{equation}
\mbox{ Then }P \{X_d =0\}/\hat {P}\{X_d =0\} \to 1 \mbox{ as } d\to \infty.
\end{equation}
\newline \noindent
Comments on conditions:
\newline\noindent
The theorem doesn't prove too much itself, but rather outlines a program for 
proving the validity of the approximation in particular cases.
\newline\noindent
Conditions I,II bound the third and fourth cumulants.
Condition III,IV require that the third and fourth cumulants affect the 
characteristic function integral through the summary cumulants $\kappa _d^3 
,\kappa _d^4 $.
Condition V requires that contributions to the characteristic function integral be negligible outside a small cube centered at 0. 
In particular this causes the determinant of the lattice of possible values of $X_d$ to be $1$ for $d$ 
large enough.
\newline\newline\noindent\textbf{ Proof:}
Let $K_{34}(t)=-{1\over 6}iK_d^3 (t)+{1 \over 72}\kappa _d^3 +{1 \over 24}K_d^4 (t)-{1 \over 24} \kappa_d^4  $.
\newline\noindent From I,II,
\begin{equation} 
\begin{array}{rcl}
E_d\{Q_\varepsilon|\exp K_{34}(t)|^2\} &\le& \exp [{1 \over 36}\kappa_d^3+{1  \over 12}K_d^4(t)-{1 \over 12}\kappa_d^4]=O(1),\\
E_d \left\{Q_\varepsilon \exp [K_{34}(t)+  o(1)]\right\}&-&E_d \left\{Q_\varepsilon \exp [K_{34}(t)]\right\}\\
&=&o(1)E_d \left\{Q_\varepsilon \exp [K_{34}(t)]\right\}\\
&=& o(1)\left (E_d\left\{Q_\varepsilon|\exp K_{34}(t)|^2\right\}\right)^{1/2}=o(1),\\
\end{array}
\end{equation}
From III,IV
\begin{equation}
\begin{array}{rcl}
&&\Delta_d(2\pi)^{-d}\int {Q_\varepsilon } (t)\phi _d(t)/\hat {P}(X_d =0\}= E_d\left\{Q_\varepsilon \exp[{1 \over 2} K^2_d(t) +\log\phi(t)+
{1 \over 72}\kappa _d^3 -{1 \over 24}\kappa _d^4 ]\right\}\\
&&=E_d \left\{Q_\varepsilon \exp [K_{34}(t)+  o(1)]\right\}=  E_d \left\{Q_\varepsilon \exp [K_{34}(t)]\right\}+o(1) \to 1 \mbox{ as } t \to \infty.
 \end{array}
\end{equation}
Thus
\begin{equation}
\int Q_\varepsilon \phi _d /\left\{(2\pi )^{d/2}| V_d | ^{-1/2}\exp  
[-{1 \over 72}\kappa _d^3 +{1 \over 24}\kappa _d^4 ]\right \} \to 1.
\end{equation}
A similar argument shows that, since $| \exp [K_d^3 (t)i^3/3!]| 
=1$,
\begin{equation}
\int Q_\varepsilon | \phi _d | /\left \{(2\pi )^{-d/2}| V_d | 
^{1/2}\exp [{1 \over 24}\kappa _d^4 ]\right \} \to 1.
\end{equation}
This shows that $\int Q_\varepsilon | \phi _d | =O(1)|\int Q_\varepsilon  \phi _d  |$.\newline
Thus from condition V ,
\begin{equation}
\int Q_\pi \phi_d   /\int Q_\varepsilon \phi_d \to 1.
\end{equation}  
\newline
\newline We now show that condition V requires the determinant $\Delta_d$ of the 
lattice to be 1 for $d$ large enough. In the contrary case, consider the 
reciprocal lattice in $d$ dimensions consisting of all vectors $a$ for which 
$a'X_d $ is integer with probability one. The determinant of this lattice is 
the reciprocal of the determinant of the original lattice, and so the 
reciprocal determinant is less than or equal to 
$\raise.5ex\hbox{$\scriptstyle 1$}\kern-.1em/ 
\kern-.15em\lower.25ex\hbox{$\scriptstyle 2$} $. There must be a non-zero 
point in the reciprocal lattice which lies in the half-unit cube; thus there is 
a  non-zero point $t=2\pi a$ lying in the cube $Q_\pi(t)=1$ for which $a'X_d$ is integer.
Now $\phi_d(t+u)=E\{\exp(i(t+u)'X_d)\}=E\{\exp(iu'X_d\}=\phi_d(u)$, since $\exp(2\pi a'X_d)=1$.
Thus the integral $| \phi _d (t)| $ in the neighbourhood of $t=2\pi a$ equals its 
integral in the neighbourhood of $0$, which contradicts V.
\newline\noindent
\newline\noindent Since $\Delta_d=1$, combining (15) and (18) gives
\begin{equation}
P\{X_d =0\}/\hat {P}\{X_d =0\}=\left(1+o(1)\right)\int Q_\pi \phi_d/\int Q_\varepsilon \phi_d  \to 1 \mbox{ as }d\to \infty ,
\end{equation}
which concludes the proof.
 
\newpage
\noindent\textbf{3 Numbers of contingency tables with given row and column sums}
\newline\newline
Consider a contingency table of non-negative integers $X_{ij} ,1\le i\le 
m,1\le j\le n$ with  row and column sums $R_i =\sum\nolimits_j 
{X_{ij}  } ,C_j =\sum\nolimits_i {X_{ij} }  $. 
 We wish to estimate the number of tables satisfying the 
constraints $R_i=r_i,C_j=c_j$. Define the dimension $d=(m+n-1)$  integer vector $S_d :$
\begin{equation}
\begin{array}{rcl}
 S_{jd} &=&R_j -r_j,1\le j\le m, \\ 
 S_{(k+m)d} &=&C_k -c_k ,1 \le k\le n-1.  
 \end{array}
\end{equation}

\noindent Following the program of section 2, the Edgeworth approximation begins with the maximum entropy distribution for
$\{X_{jk} \}$ with expectations $ER_j =r_j ,EC_k =c_k $, which consists of 
independent geometrics with expectations $\mu _{jk} $: 
\begin{equation}
P\{X_{jk} =x\}=({\mu_{jk} \over 1+\mu _{jk} })^x/(1+\mu _{jk} ),
\end{equation}
\noindent where $\log (1+1/\mu _{jk} )=\alpha _j +\beta _k $ and parameters $\alpha 
_j ,\beta _k $ are chosen for which 
\begin{equation}
ER_j=\sum\nolimits_k {\mu _{jk} } =r_j ,EC_j =\sum\nolimits_j {\mu _{jk} } =c_k. 
\end{equation}
 The existence of parameters $\alpha_j ,\beta_k$ satisfying the marginal constraints is shown in [B09]. The maximum entropy entries $\mu_{jk}$ are uniquely determined. $\alpha+c,\beta-c$ is a solution if and only if $\alpha,\beta$ is
a solution.
 
\noindent The conditional distribution of $\{X_{jk} \}$ under the constraints $\{R =r ,C =c \}$, (equivalently $\{S_d=0\}$), is uniform. 
 The number of integers satisfying the constraints is 
\begin{equation}
Q(S_d=0)=P\{S_d=0\}\exp (I(P))=P\{S_d=0\}\prod\limits_{jk} (1+\mu _{jk} 
)^{1+\mu _{jk} }\mu _{jk}^{-\mu _{jk} } .
\end{equation}
The probability$P\{S_d=0\}$ is approximated by 
\begin{equation}
\hat {P}\{S_d =0\}=(2\pi )^{-d/2}| V_d | ^{-1/2}\exp (-\kappa _d^3 
/72+\kappa _d^4 /24),
\end{equation}
 depending on the first four cumulants of $S_d $, as 
explained in section 2. See [BH09,BH10a] for further discussion.
\newline\newline\noindent
Each element of $S_d $ is the deviation from its mean of a sum of independent geometrics with 
expectations $\{\mu _{jk} \}$. The mean-centered geometric characteristic function with expectation $\mu$ is
\begin{equation}
\psi_\mu(t)=e^{-i\mu t}/(1-\mu(e^{it}-1)).
\end{equation}
From theorem 1, the validity of the asymptotic 
estimate may be assessed by the limiting behavior of the characteristic 
function of $S_d $, with parameters
\begin{equation}
\begin{array}{rcl}
&&t_j =v_j , 1\le j\le m,\quad t_{m+k} =w_k ,1\le k\le n-1,\quad w_n =0,\\
 \phi _d (t)&=&E\left\{\exp \left(i[v'(R-r)+w'(C-c)]\right)\right \}  
 =\prod\limits_{jk} \psi_{\mu_{jk}}(v_j +w_k ) 
 \end{array}
\end{equation}
We sometimes use $t$ to refer to all the parameters in the characteristic function, and at other times use $v,w$ to treat
separately the parameters in the characteristic function associated with the rows and columns respectively.

Use $x_n \sim y_n$ if $x_n/y_n \to 1$ and $x_n \approx y_n$ if $\limsup |x_n/y_n|<\infty, \limsup |y_n/x_n|<\infty.$
\newline\newline
\textbf {Theorem 2}\newline 
Suppose that , 
as $d=m+n-1 \to \infty,m \approx n \approx \min r_i \approx \max r_i \approx \min c_i \approx \max c_i.$ Assume that 
\begin{equation}
\liminf {(1+{n \over \max r_i})(1+{m \over \max c_i}) \over (1+{mn \over \sum r_i})} >1,
\end{equation}

The cumulants $K_d^r (t)$ of $t'S_d $ are the sums of the corresponding 
cumulants of the geometrics with expectations $\mu _{jk} $ and 
parameters $t_{jk} =v_j +w_k $ ,
\begin{equation}
\begin{array}{l}
 K_d^2 =\sum\nolimits_{jk} {t_{jk}^2 } \mu _{jk} (1+\mu _{jk} ) \\ 
 K_d^3 =\sum\nolimits_{jk} {t_{jk}^3 } \mu _{jk} (1+\mu _{jk} )(1+2\mu _{jk} 
) \\ 
 K_d^4 =\sum\nolimits_{jk} {t_{jk}^4 } \mu _{jk} (1+\mu _{jk} )(1+6\mu _{jk} 
(1+\mu _{jk} )) .\\ 
 \end{array}
\end{equation}

Let $K_d^2 =t'V_d t$. Let $E_d $ denote expectation with respect to $t\sim 
N(0,V_d^{-1} )$. 

Then 
\begin{equation}
P\{S_d =0\}/(2\pi )^{-d/2}| V_d | ^{-1/2}\exp (-{1 \over 72} E_d \{(K_d^3)^2\}+{1 \over 24} E_d K_d^4 )\}\to 1
\end{equation}

Remark on conditions: Our proof requires that the relative sizes of the maximum entropy entries be bounded
asymptotically,  and that the absolute sizes are bounded away from zero and infinity. In [BH09] we prove validity of the Edgeworth approximation dropping the condition that the absolute sizes be bounded away from infinity. 

\noindent {\bf Proof}:
\newline\noindent We will show conditions I-V of theorem 1 hold.

\noindent\textbf{Lemma 3.1} $\max \mu_{ij} \approx\min \mu_{ij}\approx 1$.
\newline
\newline\noindent {\bf Proof}:\newline
Let $\Omega$ be the set of $ m \times n$  matrices $\mu$ satisfying
\begin{equation}
\begin{array}{rcl}
\mu_{ij} &>& 0,\\
i \le k, j \le l &\Rightarrow& \mu_{ij} \ge \mu_{kl},\\
(1+1/\mu_{ij})(1+1/\mu_{kl})&=&(1+1/\mu_{il})(1+1/\mu_{kj}).
\end{array}
\end{equation} 
Since the previous equation holds if and only if $\log (1+1/\mu _{jk} )=\alpha _j +\beta _k $, these matrices consist of the maximum entropy geometric expectation matrices corresponding to the 
possible non-increasing positive row sums $r_1 \ge r_2 \ge ...r_m>0$ and   the 
possible non-increasing positive column sums $c_1 \ge c_2 \ge ...c_n>0$.

\noindent \textbf{Lemma 3.1.1}
\newline\noindent
The maximum entry $\mu_{11}$ achieves its maximum over $\mu \in \Omega$ for given values
of $r_1=\sum_k \mu_{1k},c_1=\sum_j \mu_{j1},T=\sum_{jk} \mu_{jk}$ when $\mu_{12}= \dots =\mu_{1j}= \dots =\mu_{1n}, \mu_{21}= \dots \mu_{i1} \dots =\mu_{m1}$.
And the minimum entry $\mu_{mn}$ achieves its minimum for given values
of $r_m,c_n,T$ when $\mu_{m1}= \dots \mu_{1j} \dots =\mu_{m(n-1)},\mu_{1n}= \dots \mu_{2n} \dots =\mu_{(m-1)n}$. 
\newline
\newline\noindent {\bf Proof}:\newline
The result is trivial if either $r_1=T/m$ or $c_1=T/n$ ; it will be useful, for uniqueness, to forbid these conditions.

We first prove that the maximum entry $\mu_{11}$ achieves its maximum over $\mu \in \Omega$ for given values
of $r_1=\sum_k \mu_{1k},c_1=\sum_j \mu_{j1},T=\sum_{jk} \mu_{jk}$ when $\mu_{12}= \dots =\mu_{1j}= \dots =\mu_{1n}, \mu_{21}= \dots \mu_{i1} \dots =\mu_{m1}$. Equivalently, since  by (30), $\mu$ is determined by its first row and column, it is equivalent to maximize $\mu_{11}$ over choices of $u=\{\mu_{j1},2 \le j \le m\},v=\{\mu_{1k},2 \le k \le n\},$ for given values of $r_1,c_1,T$. We need to show that the maximal $\mu$ occurs when $(u,z) \in \Xi$, where all the $u's$ are equal and all the $z's$ are equal.
 
Consider first the maximization of $T$ over $\mu \in \Omega$ with $r_1,c_1,\mu_{11}$ fixed, which is equivalent to  maximizing $T$ over choices of $(u,z)$ which are constrained to lie in a compact polyhedron so that $r_1,c_1,\mu_{11}$ are fixed. 
\newline\newline
Add a further constraint by fixing $z$, so that the maximization occurs by varying only the entries $u$. 
From (30), for $i>1$,
\begin{equation}
(1+1/\mu_{ij})=(1+1/\mu_{i1})(1+1/\mu_{1j})/(1+1/\mu_{11}) = \lambda_j(1+1/\mu_{i1}) 
\end{equation}
where $\lambda_j =(1+1/\mu_{1j})/(1+1/\mu_{11})\ge 1$ is fixed given the first row, and by the forbidden equality,
$\lambda_j > 1$ for at least one $j$. For $i>1$, it follows that $\mu_{ij}$ is a concave function of $\mu_{i1}$ determined by the fixed $\lambda_j$, and by the forbidden equality $\sum_j \mu_{ij}=g(\mu_{i1})$ where $g$ is strictly concave in $\mu_{i1}$, and depends only on the fixed $\lambda_j$. \newline\newline 
Thus $T=\sum_j \mu_{1j}+\sum_{i>1} g(\mu_{i1})$ is a strictly concave function of $u$ with a unique maximum at $u^0$, say. If $u^0_{i1} \ne u^0_{(i+1)1}$, then
by strict concavity of $g$,\newline $2g( {1 \over 2} [u^0_{i1}+ u^0_{(i+1)1}]) >g(u^0_{i1}) +g( u^0_{(i+1)1})$, so the function $T$ may be improved by replacing both $u^0_{i1}$ and $u^0_{(i+1)1}$ by ${1 \over 2} [u^0_{i1}+ u^0_{(i+1)1}]$, a contradiction.
Thus
$\mu_{21}= \dots =\mu_{i1}= \dots =\mu_{m1}$ at the maximum.

Now return to the  maximization of $T$ over $u,z$ with $r_1,c_1,\mu_{11}$ fixed.  The maximum of $T$, say $T(\mu_{11})$, occurs for some $(u,z)$, and it may be improved, from the previous paragraph, unless $(u,z) \in \Xi$, so these conditions hold at the maximum. In addition, the maximizing point $(u,z)$ is unique, given $r_1,c_1,T$. Thus, at the maximum,

\begin{equation}
\begin{array}{rcl}
T&=&r_1+c_1-\mu_{11}+(m-1)(n-1)\mu_{22}\\
1+{1 \over \mu_{22}}&=&(1+{m-1 \over c_1-\mu_{11}})(1+{n-1 \over r_1-\mu_{11}}){\mu_{11} \over 1+\mu_{11} }
\end{array}
\end{equation}
It will be seen from (32) that $\mu_{22}$ and therefore $T(\mu_{11})$ are both decreasing functions of $\mu_{11}$. \newline\newline
Finally, we turn to the maximization of $\mu_{11}$ over $\mu \in \Omega$ with $r_1,c_1,T=T^0$ fixed, accomplished by considering all choices of $u,z$ constrained to lie in a compact set $\Gamma$ so that $r_1,c_1,T=T^0$ are fixed.  Then $\mu_{11}=\mu_{11}^0$ is maximized at some point $(u^0,z^0)$ in $\Gamma$. If  $(u^0,z^0) \notin \Xi$, we can find,  $(u^1,z^1) \in \Xi$ maximizing $T$ for the given $r_1,c_1,\mu_{11}^0$, so that $T(\mu_{11}^0)>T^0$. Since $\mu_{11}^0$ is maximal, the value of $\mu_{11}$ at the point $(u^1,z^1)$  given $r_1,c_1,T=T^0$ must satisfy $\mu_{11}^1 \le \mu_{11}^0$. Also, the maximal value of $T$ given $r_1,c_1,\mu^1_{11}$ is achieved at the unique point $(u^1,z^1) \in \Xi$, so that $T^0=T(\mu_{11}^1)$. Since  $T(\mu_{11})$ is decreasing in $\mu_{11}$, $T^0=T(\mu_{11}^1) \ge T(\mu_{11}^0)$ which contradicts $T(\mu_{11}^0)>T^0$ and establishes that the maximum of $\mu_{11}$ over $\mu \in \Omega$ with $r_1,c_1,T=T^0$ fixed occurs at a point $(u,z) \in \Xi$ .

\noindent A similar argument is used for the minimum of $\mu_{11}$, first minimizing $T$ for fixed $r_m,c_n,\mu_{mn}$ over possible choices of the last row and column, showing that that the values of last column other than the last entry are equal, and the values of the last row other than the last entry are equal. And then transfer this result to the
minimization of $\mu_{mn}$ for fixed $r_m,c_n,T$.

This concludes the proof of lemma 3.1.1.\newline\newline

\noindent From  lemma 3.1.1, the minimum entry $\mu_{mn}$ for given $r_m,c_n,T$ occurs when
$\mu_{m1}= \dots =\mu_{1j}= \dots =\mu_{m(n-1)},\mu_{1n}= \dots =\mu_{2n}= \dots =\mu_{(m-1)n}.$
  In this case
\begin{equation}
\begin{array}{rcl}
(n-1)\mu_{m1}+\mu_{mn}=r_m \Rightarrow \mu_{m1}&=&O(1),\\
(m-1)\mu_{1n}+\mu_{mn}=c_n \Rightarrow \mu_{1n}&=&O(1),\\
(n-1)\mu_{11}+\mu_{1n}=r_1 \Rightarrow \mu_{11}&=&O(1),\\
1+1/\mu_{mn}=(1+1/\mu_{m1})(1+1/\mu_{1n})/(1+1/\mu_{11})&=&O(1).
\end{array}
\end{equation}

This guarantees that $\mu_{mn}$ is bounded away from zero in the extreme case where it takes its smallest value,
 so it must be bounded away from zero in every case. Also $\mu_{mn}<r_1/n$ is bounded away from $\infty$ by the first assumption.
Thus $\mu_{mn}\approx 1$ as required.

\noindent The maximum entry $\mu_{11}$ for given $ m,n,r_1,c_1,T$ occurs when $\mu_{12}= \dots =\mu_{1j}= \dots =\mu_{1n},\mu_{21}= \dots =\mu_{i1}= \dots =\mu_{m1}$.
We will show for this maximal entry that  $\limsup \mu_{11} < \infty$ if and only if $\liminf [(1+n/r_1)(1+m/c_1)/(1+nm/T)] > 1.$
\begin{equation}
\begin{array}{rcl}
(n-1)\mu_{1n}+\mu_{11} &=&r_1, \\
(m-1)\mu_{m1}+\mu_{11}&=&c_1, \\
(n-1)\mu_{mn}+\mu_{1n}&=&(T-r_1)/(m-1),\\
 1+1/\mu_{11}&=&(1+1/\mu_{m1})(1+1/\mu_{1n})/(1+1/\mu_{mn}).
\end{array}
\end{equation}

It follows that $\mu_{1n} \approx 1, \mu_{m1} \approx 1, \mu_{mn} \sim T/mn.$
If $\limsup \mu_{11} < \infty$, then $\mu_{1n} \sim r_1/n, \mu_{m1} \sim c_1/m,$ and
\begin{equation} 
(1+1/\mu_{11}) \sim (1+n/r_1)(1+m/c_1)/(1+nm/T) > 1.
\end{equation}
Conversely, if $\liminf [(1+n/r_1)(1+m/c_1)/(1+nm/T)] > 1,$
\begin{equation}
(1+1/\mu_{11})=(1+1/\mu_{m1})(1+1/\mu_{1n})/(1+1/\mu_{mn})>(1+n/r_1)(1+m/c_1)/(1+1/\mu_{mn}),
\end{equation}
so also $\liminf(1+1/\mu_{11})>1$, which implies $\limsup \mu_{11} < \infty$, as required.
 This concludes the proof of Lemma 3.1. 
\newline\newline\textbf {Lemma 3.3 }
\begin{equation}
 \log | V_d | -d\log d = O(d).
\end{equation}
Let $\delta_{ij} =1 \mbox{ if } i=j,\delta_{ij} =0 \mbox{ if } i\ne j.$ Then
\begin{equation}
E_d (t_{ir}t_{js})+O(d^{-2})  \approx [\delta_{ij}+\delta_{rs}]d^{-1}.
\end{equation}

{\bf Proof:} 
\newline\noindent Let $\lambda _{jk} =\mu _{jk} (1+\mu _{jk} )$. The quadratic form 
$t'V_dt=K_d^2 =\sum\nolimits_{jk} {t_{jk}^2 } \lambda _{jk} $ is increasing in 
each $\lambda _{jk}$, so that the determinant $| V_d | $ is also increasing in each 
$\lambda _{jk} $; thus  $| V_d | \le | V_d(\lambda_{11})| $ where $V_d (\lambda_{11})$ is the covariance matrix corresponding to the quadratic form $K_d^2 
=\sum\nolimits_{jk} {t_{jk}^2 } \lambda_{11}$, for which $| V_d (\lambda_{11})| 
=\lambda_{11}^dm^{n-1}n^{m-1}$. Similarly, $| V_d | \ge | V_d (\lambda_{mn})| =\lambda_{mn}^dm^{n-1}n^{m-1}$.
Thus $\log | V_d | -d\log d =O(d)$. This result may also 
be obtained by noting that $| V_d | $ is a sum of $m^{n-1}n^{m-1}$products of $d$ coefficients 
$\lambda _{jk} $.

Again, since the quadratic form $t'V_dt$ is increasing in each $\lambda _{jk} 
$, necessarily the quadratic form $t'V_d^{-1}t$ is decreasing in each $\lambda 
_{jk} $, so  bounds for the variances induced by $t\sim 
N(0,V_d^{-1} )$ are obtained by setting all the $\lambda _{jk} $ equal to 
$\lambda_{11}$ or to $\lambda_{mn}$. This establishes that $E_d t_i^2 \approx d^{-1}$.

To bound the off-diagonal terms in $V_d$, note that $t\sim N(0,V_d^{-1} )$ allows 
us to determine the conditional distribution $v| w$ from the quadratic 
form $t'V^{-1}t$ with $w$ fixed, and similarly the conditional distribution 
$w| v$. Indeed the $v_j $ are independent given $w$, and the $w_k $ are 
independent given $v$. This gives a relationship between the $v$ and $w$ 
covariance matrices which produces the required bound on the 
off-diagonal terms. A result similar to lemma 3.3 is proved in [BH09] using non-probabilistic 
methods.

\begin{equation}
\begin{array}{l}
 \mbox{ Define } \alpha 
=1/\sum\nolimits_{ij} {\lambda _{ij} },\quad \alpha _i =1/\sum\nolimits_j {\lambda _{ij} } ,\quad \alpha _{ij} =\alpha _i \lambda _{ij} ,\\
\bar {v}=\alpha \sum\nolimits_{ij} {\lambda _{ij} v_i },\quad 
\bar {w}=\alpha \sum\nolimits_{ij} {\lambda _{ij} w_j },\quad\tilde {v}_i =v_i -\bar {v},\quad\tilde {w}_j =w_j -\bar {w} .\\  
\end{array}
\end{equation} 
 
 \noindent Note that $\mu_{11}  \approx \mu_{mn} \approx 1 \Rightarrow \min _{ij} (\alpha _{ij} /\alpha 
\sum\nolimits_k {\lambda _{kj} )-\varepsilon \ge 0}$ some $\varepsilon >0.$ 
 \newline And $v_i | w\sim N(-\sum\nolimits_r {\alpha _{ij} w_j } ,\alpha _i 
)$ independently for different $i$. Then
\begin{equation}
\begin{array}{rcl} 
 E\{\bar {v }| w\}&=&E\{\sum\nolimits_i ( \alpha /\alpha _i )v _i| w\}=-\sum\nolimits_i \alpha \sum\nolimits_r {\lambda _{ij} } w_j 
=-\bar {w}, \\ 
 E\{\tilde {v }_i | w\}&=&-\sum\nolimits_j {\alpha _{ij} } \tilde {w}_j, 
\\ 
 E_d \{v_i v_j | w\}&=&\alpha _i \delta _{ij} +\sum\nolimits_{rs} {\alpha 
_{ir} \alpha _{js} } w_r w_s , \\ 
 E_d \{\tilde {v}_i v_j | w\}&=&\alpha _i \delta _{ij} -\alpha 
+\sum\nolimits_{rs} {\alpha _{ir} } \tilde {w}_r \alpha _{js} w_s ,\\ 
 E_d \{\tilde {v }_i \bar {v}| w\}&=&\sum\nolimits_{rs} {\alpha _{ir} } 
\tilde {w}_r \bar {w} ,\\ 
 E_d \{\tilde {v}_i \tilde {v}_j \}&=&\alpha _j \delta _{ij} -\alpha 
+\sum\nolimits_{rs} {\alpha _{ir} } \alpha _{js} E_d \{\tilde {w}_r \tilde 
{w}_s \}, \\ 
 &=&\alpha_j  \delta _{ij} -\alpha +\sum\nolimits_{rs} (\alpha _{ir}  -\varepsilon \alpha \sum\nolimits_k \lambda _{k
r}  
)\alpha _{js} E_d\{\tilde {w}_r \tilde {w}_s \}\mbox{ since }\sum\nolimits_{kr} {\lambda _{kr} } \tilde {w}_r =0. \\ 
\end{array}
\end{equation}
Note that $E_d t_i^2 \approx d^{-1}\Rightarrow $  $E_d \tilde {w}_r^2 =O(d^{-1}).\newline
$ Also $\alpha \approx d^{-2},\max \alpha _{ij}\approx d^{-1},\tilde {\alpha }_{ir} 
=\alpha _{ir} -\varepsilon \alpha \sum\nolimits_k {\lambda _{kr} } \ge 
0,\sum\nolimits_r {\tilde {\alpha }_{ir} =1-} \varepsilon.$
 
\begin{equation}
\begin{array}{rcl}
 E_d \tilde {v}_i \tilde {v}_j &\le& \alpha _j \delta _{ij} -\alpha +O(d^{-1})\max 
(E_d \tilde {w}_r^2 )+\sum\nolimits_{r\ne s} {\tilde {\alpha }_{ir} } \alpha 
_{js} \max _{r\ne s} E_d \tilde {w}_r \tilde {w}_s, \\ 
 &\le& \alpha _j \delta _{ij} +O(d^{-2})+(1-\varepsilon )\max _{r\ne s} E_d \tilde 
{w}_r \tilde {w}_s, \\ 
 \max _{i\ne j} E_d \tilde {v}_i \tilde {v}_j &\le& O(d^{-2})+(1-\varepsilon )\max 
_{r\ne s} E_d \tilde {w}_r \tilde {w}_s. \\ 
\end{array}
 \end{equation}

\noindent Similarly, 
\begin{equation}
\begin{array}{rcl}
\min _{i\ne j} E_d \tilde {v}_i \tilde {v}_j &\ge& 
O(d^{-2})+(1-\varepsilon )\min _{r\ne s} E_d \tilde {w}_r \tilde {w}_s ,\\
\max _{i\ne j} | E_d \tilde {v}_i \tilde {v}_j | &\le& 
O(d^{-2})+(1-\varepsilon )\max _{r,s} | E_d \tilde {w}_r \tilde {w}_s |. 
\end{array}
 \end{equation}
The joint distribution of the $\tilde {v}_i ,\tilde {w}_r $ depends on the 
joint distribution of the $t_{jk} $ and so does not depend on the particular 
particular linear combination of $v_j ,1\le j\le m_, w_k ,1\le k\le n$ that 
is set zero to reduce the dimensionality of these $m+n$ terms to $d=(m+n-1)$. Thus the reverse 
result holds conditioning on the $\tilde {v}_i $:
\begin{equation}
\begin{array}{rcl}
 \max _{r \ne s} | E_d \tilde {w}_r \tilde {w}_s | &\le&O(d^{-2})+(1-\varepsilon )\max _{i\ne j} | E_d \tilde {v}_i \tilde {v}_j| \\ 
 \max_{i\ne j} | E_d \tilde {v}_i \tilde {v}_j | ,\max _{r\ne s} | E_d \tilde {w}_r \tilde {w}_s | &=& O(d^{-2}) \\ 
 \end{array}
 \end{equation}

\noindent A similar argument shows that $\mathop {\max }\limits_{i,j} | E_d \tilde 
{w}_i \tilde {v}_j | =O(d^{-2})$. Also
\begin{equation}
t'Vt=\sum\nolimits_{ij} {t_{ij}^2 } \lambda _{ij} =\sum\nolimits_{ij} 
(\tilde {v}_i  +\tilde {w}_j )^2\lambda _{ij} +(\bar {v}+\bar 
{w})^2/\alpha 
\end{equation}
so that $\bar {v}+\bar {w}$ is independent of $\tilde {v}_i ,\tilde {w}_j $ 
with variance $\alpha \approx d^{-2}$.
Concluding the proof of lemma 3.3,
\begin{equation}
\begin{array}{rcl} 
 E_d (t_{ir}t_{js}) &=& E_d (\tilde {v}_i +\tilde {w}_r +\bar {v}+\bar {w})(\tilde {v}_j +\tilde {w}_s +\bar {v}+\bar{w}) \\ 
 &=& E_d \tilde {v}_i \tilde {v}_j +E_d \tilde {v}_i \tilde {w}_s +E_d\tilde {w}_r \tilde {v}_j +E_d \tilde {w}_r \tilde {w}_s +E_d (\bar {v}+\bar {w})^2  \\ 
E_d (t_{ir}t_{js})+O(d^{-2})& \approx& [\delta_{ij}+\delta_{rs}]d^{-1}.
\end{array}
\end{equation}
 
We now apply theorem 1 by verifying the conditions I-V. Similar propositions to I-IV are proved using similar methods in [BH09].
\newline\newline\noindent {\bf CONDITION I:} $\kappa _d^3 =O(1) ,\kappa _d^4 =O(1).$
\begin{equation}
\begin{array}{rcl}
\kappa _d^3 =E_d (K_d^3 )^2&=&E_d (\sum\nolimits_{jk} {t_{jk}^3 } \mu _{jk} 
(1+\mu _{jk} )(1+2\mu _{jk} ))^2=O(\sum\nolimits_{jkrs} |E_d t_{jk}^3 
t_{rs}^3)|,\\
E_d t_{jk}^3 t_{rs}^3 &=&9E_d t_{jk}^2 E_d t_{rs}^2 E_d t_{jk} t_{rs} +6(E_d 
t_{jk} t_{rs} )^3
\end{array}
\end{equation}
From lemma 3.3, 
\begin{equation}
\begin{array}{rcl}
E_d t_{jk} t_{rs} &=&O(d^{-2}+(\delta _{jr} +\delta _{ks})d^{-1}), \\
E_d t_{jk}^3 t_{rs}^3 &=& O(d^{-4}+(\delta _{jr} +\delta _{ks})d^{-3}).
\end{array}
\end{equation}
In the $O(d^4)$ terms in the sum $\sum\nolimits_{jkrs} {E_d t_{jk}^3} 
t_{rs}^3$, there are $O(d^3)$ terms in which $(\delta _{jr} +\delta_{ks} )>0$; thus the sum over all terms is $O(1)$.
\newline\newline\noindent $\kappa _d^4 =E_d K_d^4 $ is the sum of $d^2$ terms of $O(d^{-2})$, 
so it also is bounded.
\newline\newline\noindent{\bf CONDITION II:} $E_d \{Q_\varepsilon \exp [{1 \over 12} K_d^4 (t) ]\}=O(1).$

\noindent For $X,Y$ joint normal with mean zero, 
\begin{equation}
\begin{array}{rcl}
\mbox{cov}(X^4,Y^4)&=&72EX^2 EY^2 E^2XY+24E^4XY\\
\mbox{cov}(t_{jk}^4 ,t_{rs}^4 )&=&O(d^{-6}+(\delta _{jr} +\delta _{ks} )d^{-4})
\end{array}
\end{equation}
\noindent Since there are only $d^3$ covariances for which $(\delta _{jr} +\delta 
_{ks} )>0$,
\begin{equation}
E_d(K_d^4-\kappa_d^4)^2=E_d (K_d^4 -E_d K_d^4 )^2=O(\sum\nolimits_{jkrs} |\mbox{cov}(t_{jk}^4 ,t_{rs}^4 )|)=O(d^{-1}).
\end{equation}

\noindent From [D87] Corollary 5, 
since $K_d^4 -\kappa_d^4 $ is a polynomial of degree 4 in Gaussian variables,
\begin{equation}
\begin{array}{rcl}
r>1\Rightarrow E_d | K_d^4 -\kappa_d^4 | ^{2r}&\le& r^{4r}[E_d (K_d^4-\kappa_d^4 )]^r\le C_r d^{-r},\\
P_d \{K_d^4 \ge \kappa_d^4 +1\}&\le& C_r d^{-r}.
\end{array}
\end{equation}
\noindent When $t\sim N(0,V_d^{-1} )$, the multivariate normal density is $A\exp 
[-\textstyle{1 \over 2}K_d^2 (t)].$ Thus $E_d \exp [\alpha K_d^2 (t)]=(1-2\alpha )^{-d}.$
Also, since the $\mu _{jk} $ are bounded, $K_d^4 Q_\varepsilon\le C\varepsilon 
^2K_d^2Q_\varepsilon $. Thus

\begin{equation}
\begin{array}{rcl}
 E_d Q_\varepsilon \exp [{1 \over 12} K_d^4 (t) ]&\le& E_d \exp [{1 \over 12}(\kappa _d^4 
+1)]+E_d \{K_4^d \ge \kappa _d^4 +1\}\exp ({1 \over 12}C\varepsilon^2 K_d^2 ) \\ 
 &=&O(1)+{E_d}^{1/2} \{K_4^d >\kappa _d^4 +1\}\mbox{ }(1-{1 \over 6}CM^2\log d/d)^{-d/2} \\ 
 &=&O(1)+C_r^{1/2} d^{-r/2}d^{{1 \over 3}CM^2}=O(1) \mbox{ for } r>CM^2 .\\ 
\end{array}
\end{equation}
\newline\newline \noindent {\bf CONDITION III: }$ Q_\varepsilon[\log \phi _d (t)-\sum\nolimits_{r=2}^4 {K_d^r } (t){\textstyle{i^r} \over {r!}}]=o(1) 
  $.
\newline\noindent For a geometric with mean $\mu \approx 1$, the log centered 
characteristic function $\psi_\mu$ has the standard Taylor series expansion:
\begin{equation}
\begin{array}{rcl}
\log \psi _\mu (t)&=&\sum\nolimits_{r=2}^4 {K^r(it)^r} /r!+O(1)| t| ^5,\\
\log \phi _d (t)&=&\sum\nolimits_{jk} {\log \psi _{\mu _{jk} } (t_{jk} )=} 
\sum\nolimits_{r=2}^4 {K_d^r i^r} /r!+O(1)\sum\nolimits_{jk} {| t_{jk} 
| } ^5,\\
\sum\nolimits_{jk} {| t_{jk} } |^5 &=& O(d^2 \varepsilon^5)= o(1), 
\end{array}
\end{equation}
as required.
\newline\newline\noindent{\bf CONDITION IV:}
$E_d Q_\varepsilon \exp [-iK_d^3 (t)/6+\kappa _d^3 /72+K_d^4 (t)/24-\kappa _d^4 /24\}]\to 1.$
\newline\noindent We will first show that $K_d^3 =\sum\nolimits_{jk} {t_{jk}^3 } \mu _{jk} (1+\mu 
_{jk} )(1+2\mu _{jk} )$ has the same moments in the limit as a normal 
distribution $N(0,\kappa _d^3 )$. 

Define $u_\alpha =t_{jk} [\mu _{jk} (1+\mu _{jk} )(1+2\mu _{jk} )]^{1/3}$ 
where $\alpha$ ranges over the pairs of indices in $A=\{(j,k),1\le j\le m,1\le k\le 
n\}$. Let $G$ be the graph on $A$ with edges $(\alpha,\beta) \in G$ whenever either the first or second index of
$\alpha,\beta$ are the same. In particular, $(\alpha,\alpha) \in G$.
\begin{equation}
\begin{array}{rcl}
 E_d u_\alpha u_\beta &=&O(d^{-2+G(\alpha ,\beta)}),\\
 E_d u_\alpha^3 u_\beta^3 &=&O(d^{-4+G(\alpha ,\beta)}).\\
\end{array}
\end{equation}

Let $\{X_\alpha \}$ denote a multivariate normal with $EX_\alpha =0,EX_\alpha X_\beta =E_d u_\alpha ^3 u_\beta ^3 $.
We will show that $K_d^3 =\sum\nolimits_\alpha {u_\alpha ^3 } $ and $\sum\nolimits_\alpha {X_\alpha } $ have moments 
differing by $O(d^{-1})$. The first two moments are identical, by definition, and the odd 
moments are zero for both variables. For the $2r$th  moment:

\begin{equation}
\begin{array}{rcl}
E_d (\sum\nolimits_\alpha {u_\alpha ^3 } )^{2r}&=&\sum\nolimits_\alpha {E_d (u_{\alpha_1}^3 u_{\alpha_2} ^3 }.. u_{\alpha_{2r}}^3 
)\quad, \\
E(\sum\nolimits_\alpha {X_\alpha } )^{2r}&=&\sum\nolimits_{\alpha } E(X_{\alpha_1} X_{\alpha_2}.. X_{\alpha_{2r}}).
\end{array}
\end{equation}

The terms $E_d (u_{\alpha_1}^3 u_{\alpha_2} ^3 .. u_{\alpha_{2r}}^3$
tend to be larger when many of the pairs  of
$\alpha_i$ have edges in $G$; this size is compensated by the fact that fewer sets of $\alpha_1,\alpha_2,..\alpha_{2r}$
have many edged pairs. In order to count such sets, for each $\alpha_1,\alpha_2,..\alpha_{2r}$ we define a set of 
directed trees $\tau(\alpha)=\{\tau_1(\alpha),\tau_2(\alpha),..\tau_t(\alpha)\}$ on the $\alpha$-indices $(1,2,..,2r)$.

 The tree $\tau_1(\alpha)$ is initialised with root $1$; then progress through the $\alpha$-indices in order,  attaching $j$ to $k$ if $(\alpha_j,\alpha_k) \in G$ , and $k$ is the smallest index already attached to the tree for which $(\alpha_j,\alpha_k) \in G$.
The tree $\tau_i(\alpha)$ is constructed similarly on the set of $\alpha$-indices not attached to the trees 
$\{\tau_1(\alpha),\tau_2(\alpha),..\tau_{i-1}(\alpha)\}$; begin with the root $r_i$, the lowest $\alpha$-index not attached to previous trees, and progress through the $\alpha$-indices in order,  attaching $j$ to $k$ if $(\alpha_j,\alpha_k) \in G$ , and $k$ is the smallest $\alpha$-index already attached to the tree $\tau_i(\alpha)$for which $(\alpha_j,\alpha_k) \in G$.

For a set of trees $\tau=\{\tau_1,\tau_2,..\tau_t\}$ partitioning the $\alpha$-indices $\{1,..2r\}$, the number of $\alpha_1,\alpha_2,..\alpha_{2r}$ for which $\tau(\alpha)=\tau$ is
$O(d^{2r+t})$; to see this, consider the $i$th tree $\tau_i$ which has ,say, $\alpha$-indices $j_1=r_i,j_2,..j_{n_i}$ .
As $\alpha_{j_1},\alpha_{j_2},..\alpha_{j_{n_i}}$ pass through the
$O(d^{2n_i})$ possible values in $A^{n_i}$, $\alpha_{j_1}$ passes through $O(d^2)$ values, but the remaining $\alpha_{j_k}$ in the tree $\tau_i$ each pass through only $O(d)$ values, since each such $\alpha_{j_k}$ is constrained by 
$(\alpha_{j_k},\alpha_{j_{k'}}) \in G$ for some fixed $k' <k$.
 Thus the number of $\alpha_{j_1},\alpha_{j_2},..\alpha_{j_{n_i}}$ with $\tau_i(\alpha)=\tau_i$ is $O(d^{n_i+1})$.
 Noting that $\sum_i n_i=2r$, the number of $\alpha_1,\alpha_2,..\alpha_{2r}$ for which $\tau(\alpha)=\tau$ is the product of these quantities $O(d^{2r+t})$.

For a particular $\alpha_1,\alpha_2,..\alpha_{2r}$ with trees $\tau(\alpha)$ of sizes $n_1,..n_t$ , Wick's formula for $E(X_{\alpha_1} X_{\alpha_2}.. X_{\alpha_{2r}})$
is the sum over all partitions into $r$ sets of pairs of variables, of the product of the covariances for those variables. The maximal order products
occur when the pairs of variables designated in the partition lie as frequently as possible within one of the trees in $\tau(\alpha)$. Smaller order terms may be ignored because their number is bounded for $r$ fixed. If all the tree sizes $n_i$ are even, the maximal product occurs when each pair of variables designated in the partition has an edge in one of the trees; the covariance for each such variable is $O(d^{-3})$, so the product is $O(d^{-3r})$. If there are $s$ odd terms $n_i$, there are $s/2$ pairs lying in different trees and having smaller covariances, so 
\begin{equation}
\begin{array}{rcl}
E(X_{\alpha_1} X_{\alpha_2}.. X_{\alpha_{2r}})&=&O(d^{-3r-s/2}),\\
\sum_{\alpha | \tau(\alpha)=\tau}E(X_{\alpha_1} X_{\alpha_2}.. X_{\alpha_{2r}})&=&O(d^{-r+t-s/2}).
\end{array}
\end{equation}

Now 
\begin{equation}
-r+t-s/2=-\frac{1}{2}\sum n_i +t -s/2=\frac {1}{2} \sum_{n_i \mbox{ even}}(2-n_i) + \frac{1}{2}\sum_{n_i \mbox{ odd}}(1-n_i).
\end{equation}
Thus 
\begin{equation}
\begin{array}{rcl}
\sum_{\alpha | \tau(\alpha)=\tau}E(X_{\alpha_1} X_{\alpha_2}.. X_{\alpha_{2r}})&=&O(1) \mbox{ if } \max n_i \le 2,\\
\sum_{\alpha | \tau(\alpha)=\tau}E(X_{\alpha_1} X_{\alpha_2}.. X_{\alpha_{2r}})&=&O(d^{-1}) \mbox{ if } \max n_i > 2.
\end{array}
\end{equation}

For a particular $\alpha_1,\alpha_2,..\alpha_{2r}$ with  trees $\tau(\alpha)$ of sizes $n_1,..n_t$, Wick's formula for 
$E_d (u_{\alpha_1}^3 u_{\alpha_2}^3 .. u_{\alpha_{2r}}^3) $
 is the summation over all partitions into $3r$ sets of pairs of variables, of the product of the covariances for those variables. Again, the maximal terms occur when the pairs of variables lie as frequently as possible within the trees of $\tau$. If all the tree sizes are even, the maximal product is $O(d^{-3r})$. If there are $s$ odd tree sizes, there are $s/2$ pairs with smaller covariances, so again

\begin{equation}
\begin{array}{rcl}
\sum_{\alpha | \tau(\alpha)=\tau}E_d (u_{\alpha_1}^3 u_{\alpha_2} ^3 .. u_{\alpha_{2r}}^3)&=&O(1) \mbox{ if } \max n_i \le 2,\\
\sum_{\alpha | \tau(\alpha)=\tau}E_d (u_{\alpha_1}^3 u_{\alpha_2} ^3 .. u_{\alpha_{2r}}^3)&=&O(d^{-1}) \mbox{ if } \max n_i > 2.
\end{array}
\end{equation}

Thus, in equation (54) we need only consider summation over $\alpha$ whose trees have maximal size $2$. 
Let $\tau(2k,2r-2k)$ denote the trees $\{(1),(2),(3),..(2k)(2k+1,2k+2)...(2r-1,2r)$. There are ${2r \choose 2k}$ such $\tau$ with $2k$ elements of size 1 and $r-k$ elements of size 2.

{\bf Case 1. $\rho(0,2r)$: All trees of size 2}

For example, $\alpha=(11),(12),(22),(23)$ has trees $\{(1,2),(3,4)\}$.

For a particular $\alpha$ with this partition,
Wick's formula for $E(X_{\alpha_1} X_{\alpha_2}.. X_{\alpha_{2r}})$ gives a term $E(X_{\alpha_1} X_{\alpha_2}).. E(X_{\alpha_{2r-1}}X_{\alpha_{2r}})$ of $O(d^{-3r})$ when the Wick's  partition corresponds to $\tau(0,2r)$, and terms of $O(d^{-3r-1})$ when the Wick's partition includes some terms that are not concordant with $\tau(0,2r)$.
\newline\noindent
Also, Wick's formula for $E_d (u_{\alpha_1}^3 u_{\alpha_2}^3 .. u_{\alpha_{2r}}^3)$ 
gives a term $E(u^3_{\alpha_1} u^3_{\alpha_2}).. E(u^3_{\alpha_{2r-1}}u^3_{\alpha_{2r}})$ 
of $O(d^{-3r})$ by summing over the partitions of the $6r$ variables $u_{\alpha_i}$ that conform to $\tau(0,2r)$; 
for example, the variables $u_{\alpha_1},u_{\alpha_1},u_{\alpha_1},u_{\alpha_2},u_{\alpha_2},u_{\alpha_2}$ will be paired in 15 ways. All other partitions of the $6r$ variables have at least one pairing not conforming with $\tau(0,2r)$, and the corresponding covariance for that pair is $O(d^{-2})$, so that the contribution of all other partitions is $O(d^{-3r-1})$.

By definition,$E(X_{\alpha_1} X_{\alpha_2}).. E(X_{\alpha_{2r-1}}X_{\alpha_{2r}})=E(u^3_{\alpha_1} u^3_{\alpha_2}).. E(u^3_{\alpha_{2r-1}}u^3_{\alpha_{2r}})$.  Thus

\begin{equation}
 E(X_{\alpha_1} X_{\alpha_2}.. X_{\alpha_{2r}}) =E(u^3_{\alpha_1} u^3_{\alpha_2}.. u^3_{\alpha_{2r-1}}u^3_{\alpha_{2r}})+O(d^{-3r-1}).
\end{equation}

\noindent {\bf Case 2 $\rho(2r,0)$: All trees of size 1.}

For a particular $\alpha$ with this partition,
 Wick's formula for $E(X_{\alpha_1} X_{\alpha_2}.. X_{\alpha_{2r}})$
sums $E(X_{\alpha_{i_1}} X_{\alpha_{i_2}}).. E(X_{\alpha_{i_{2r-1}}}X_{\alpha_{i_{2r}}})$ 
over all partitions of $\alpha$ into $r$ pairs of variables.  Wick's formula for $E_d (u_{\alpha_1}^3 u_{\alpha_2}^3 .. u_{\alpha_{2r}}^3)$ 
consists of a leading term in which, for each $i$,  two of the $u_{\alpha_i}$ are paired; 
the other terms have at least one $u_{\alpha_i}$ 
paired with three $u_\alpha$'s that it is unlinked to, and the corresponding covariances have smaller order. The leading term is thus the sum
$9^{r}Eu^2_{\alpha_1}Eu^2_{\alpha_2}...Eu^2_{\alpha_{2r}}E(u_{\alpha_{i_1}} u_{\alpha_{i_2}}).. E(u_{\alpha_{i_{2r-1}}}u_{\alpha_{i_{2r}}})$ 
over all partitions of $\alpha$ into $r$ pairs of variables.

Noting that $E(X_{\alpha_{i_1}} X_{\alpha_{i_2}})=9Eu^2_{\alpha_1}Eu^2_{\alpha_2}E(u_{\alpha_1}u_{\alpha_2})+O(d^{-6})$, obtain that
\
\begin{equation}
 E(X_{\alpha_1} X_{\alpha_2}.. X_{\alpha_{2r}}) =E(u^3_{\alpha_1} u^3_{\alpha_2}.. u^3_{\alpha_{2r-1}}u^3_{\alpha_{2r}})+O(d^{-4r-1}).
\end{equation}

{\bf Case 3 $\rho(2k,2r-2k)$: $2k$ trees of size 1, $r-k$ trees of size 2}

For a particular $\alpha$ with this tree,
Wick's formula for $E(X_{\alpha_1} X_{\alpha_2}.. X_{\alpha_{2r}})$ has \newline leading product terms in which the partition of the $2r$ terms is such that the
terms $X_{\alpha_{2k+1}} X_{\alpha_{2k+2}}.. X_{\alpha_{2r}}$ are paired conforming to the last $r-k$ trees of size $2$ in $\tau(2k,2r-2k)$. Thus
\begin{equation}
E(X_{\alpha_1} X_{\alpha_2}.. X_{\alpha_{2r}})=
E(X_{\alpha_1} X_{\alpha_2}.. X_{\alpha_{2k}}) E(X_{\alpha_{2k+1}} X_{\alpha_{2k+2}}.. X_{\alpha_{2r}})+O(d^{-3r-k-1})
\end{equation}
Similarly, for a particular $\alpha$ with this partition,
Wick's formula for $E_d (u_{\alpha_1}^3 u_{\alpha_2}^3 .. u_{\alpha_{2r}}^3)$  has leading terms in which the partition of the $6r$ terms is such that the
terms $u^3_{\alpha_{2k+1}} u^3_{\alpha_{2k+2}}.. u^3_{\alpha_{2r}}$ are paired conforming the last $r-k$ trees of size $2$ in $\tau(2k,2r-2k)$. Thus

\begin{equation}
E(u^3_{\alpha_1} u^3_{\alpha_2}.. u^3_{\alpha_{2r}})=
E(u^3_{\alpha_1} u^3_{\alpha_2}.. u^3_{\alpha_{2k}}) E(u^3_{\alpha_{2k+1}} u^3_{\alpha_{2k+2}}.. u^3_{\alpha_{2r-1}}u^3_{\alpha_{2r}})+O(d^{-3r-k-1})
\end{equation}

From the equivalences in case 1 and case 2,
\begin{equation}
E(u^3_{\alpha_1} u^3_{\alpha_2}.. u^3_{\alpha_{2r}})=E(X_{\alpha_1} X_{\alpha_2}.. X_{\alpha_{2r}})+O(d^{-3r-k-1})
\end{equation}
Since there are $O(d^{3r+k})$ different $\alpha$ with the trees $\tau(2k,2r-2k)$,
\begin{equation}
\sum_{\tau(\alpha)=\tau(2k,2r-2k)}E(u^3_{\alpha_1} u^3_{\alpha_2}.. u^3_{\alpha_{2r}})=\sum_{\tau(\alpha)=\tau(2k,2r-2k)}E(X_{\alpha_1} X_{\alpha_2}.. X_{\alpha_{2r}})+O(d^{-1})
\end{equation}
Since this equivalence holds for all partitions with element size at most 2, and the contributions from other partitions are negligible,
\begin{equation}
\sum E(u^3_{\alpha_1} u^3_{\alpha_2}.. u^3_{\alpha_{2r}})=\sum E(X_{\alpha_1} X_{\alpha_2}.. X_{\alpha_{2r}})+O(d^{-1})
\end{equation}
as required.

We have shown that $K_d^3 =\sum\nolimits_\alpha {u_\alpha ^3 } $ and $\sum\nolimits_\alpha {X_\alpha } $ have moments 
differing by $O(d^{-1})$.
Since $\sum_\alpha X_\alpha (\kappa _d^3 )^{-1/2}\sim N(0,1)$, and a 
normal random variable is determined uniquely by its moments, $K_d^3 (\kappa 
_d^3 )^{-1/2}\to N(0,1) \mbox{ in distribution as } d\to \infty .$
\newline\newline\noindent For $Z\sim N(0,1)$, $P\{| Z| >A\}\le \exp (-{1 \over 2}A^2)$.\newline 
Thus $Q_\varepsilon \to 1 \mbox{ in probability as } d\to \infty $, since for, $M$ large enough 
\begin{equation} 
P_d\{Q_\varepsilon =0\}\le \sum_i P_d  \{| t_i | >M\sqrt {\log d/d} \}\le d\exp (-M^2\log d/O(1))\to 0 \mbox{ as } d\to \infty .
\end{equation}

Thus $E_d Q_\varepsilon \exp [-iK_d^3 (t)T (\kappa _d^3 )^{-{1 \over 2}}]\to 
E\exp [iT N(0,1)]=\exp (-{1 \over 2}T ^2)$
uniformly in any finite interval $T^2\le A$. Since $| \kappa _d^3 | \le C$, the convergence is uniform in $T ^2\le \kappa_d^3A/C$.
Now choose $A=C$ to get convergence at $T=(\kappa^3_d)^{-1/2}:$
\begin{equation}
E_d Q_\varepsilon \exp [-{1 \over 6}iK_d^3 (t)+{1 \over 72}\kappa _d^3]\to 1.
\end{equation}

Since $K_d^4 -\kappa _d^4 \to 0$ in probability, and using condition IV,
\begin{equation}
\begin{array}{rcl}
 &&| E_d Q_\varepsilon \exp [-{1 \over 6}iK_d^3 (t)+{1 \over 72}\kappa _d^3 ] (\exp {1 \over 24}[K_d^4 
(t)-\kappa _d^4 ]-1)| \\ 
 &\le& CE_d Q_\varepsilon | \exp {1 \over 24}[K_d^4 -\kappa _d^4 ]-1| \to 
0 
\end{array} 
 \end{equation}
Thus, as required, $E_d Q_\varepsilon \exp [-{1 \over 6}iK_d^3 (t)+{1 \over 72}\kappa _d^3 
+{1 \over 24}K_d^4 (t)-{1 \over 24} \kappa _d^4 ]\to 1.$
\newline\newline
\noindent{\bf CONDITION V: }
\noindent For some $ M$, $\varepsilon =M\sqrt {\log d/d} , 
\int_ {Q_\pi -Q_\varepsilon } \vert \phi _d \vert /\int_ {Q_\varepsilon} |\phi _d |=o(1).$
\newline A similar result is proved in [BH09] using analytic methods.
\newline\newline\noindent{\bf Proof:}
We define a probability $P_d$ on $t_1,...t_{m+n}=v_1,..v_m,w_1,..w_n \in (-\pi,\pi]^{m+n}$  with density proportional to $|\phi_d|$.  
To prove condition V, we need to show, for some $M$, $P_d\{\max_i|t_i|\ge \varepsilon|t_{m+n}=0\} \to 0 \mbox{ as } d \to \infty $.  
The method evaluates the conditional probability of large deviations in any single parameter $t_i$ when the rest of the parameters are well behaved. 

Since the geometric variable is integer, the geometric characteristic function has period $2\pi, $ so individual geometric characteristic functions $\psi_{\mu_{jk}}$ have values near $1$ when the argument $v_j+w_k$ has values near $2\pi$ or $-2\pi$. This will not happen for many pairs $v_j,w_k$, but is best handled by  transformation of each $v_j$ and $w_k$ from $(-\pi ,\pi ]$ to the unit 
circle $\{x| e^{ix}=1\}$: 
\begin{equation}
 \tilde {v}_j =e^{-iv_j },\tilde {w}_k 
=e^{iw_k },\bar {v}=\textstyle{1 \over m}\sum\nolimits_j {\tilde {v}_j },\bar {w}=\textstyle{1 \over n}\sum\nolimits_k {\tilde {w}_k } .
\end{equation}
 \newline
\newline\noindent{\bf Lemma 3.4:} With constants $O(1)$ independent of $d,j,k$,
\begin{equation}
 \exp [-| \tilde {v}_j -\tilde {w}_k |^2 O(1)]\le | \psi_{\mu_{jk}}(v_j+w_k) | \le  \exp [-| \tilde v_j -\tilde w_k |^2/O(1)].
\end{equation}

\noindent{\bf Proof:}
\newline For constants $k(\mu),K(\mu)$, and for all $t$,
\begin{equation}
\exp[-|e^{it}-1|^2 k(\mu)] \le |\psi_\mu(t)|^2={1 \over 1+\mu(\mu+1)|e^{it}-1|^2} \le \exp[-|e^{it}-1|^2 K(\mu)].\\
\end{equation}

Also $| e^{i(v_j +w_k )}-1| ^2= | \tilde {v}_j -\tilde {w}_k |^2$. Since $\mu_{jk} \approx 1$, the lemma is proved.
\newline
\newline\noindent {\bf Lemma 3.5 }:
\begin{equation}
\begin{array}{rcl}
 \mbox{ Define } R^2&=&\sum_{jk}| \tilde v_j -\tilde w_k |^2.\\
\mbox{ Then, for some } M , P_d\{R>d\varepsilon\}&=&\exp[-d/O(1)].\\
\end{array}
\end{equation}

This lemma guarantees that only $t$ values where most of the $| \tilde v_j -\tilde w_k |$ are small make significant contributions to the 
probabilities $P_d$.
\newline\newline\noindent {\bf Proof }:
From (70), 
\begin{equation}
 \prod_{jk} |\psi_{\mu_{jk}}(v_j+w_k)|\le \exp[-R^2/O(1)].
\end{equation}
We have previously used $\int$ to denote integration over the $d$ variables $t_1,..t_{m+n-1}$, and we will now use 
$\int_{m+n}$ to denote integration over all variables $t_1,..t_{m+n}$.
From conditions I-IV, theorem 2 implies that $\int Q_\varepsilon \phi _d /\hat {P}\{X_d =0\}\to 1$, \newline
 so $\int | \phi _d | \ge |\int Q_\varepsilon| \phi _d |=\exp(-{1 \over 2} d\log d +O(d))$ .
 The integral of $|\phi_d|$ over the first $m+n-1$ parameters is the same for each choice of $t_{m+n}$, so the integral over all $m+n$ parameters is $\int_{m+n}|\phi_d|=2\pi \int |\phi_d|.$
\newline\newline\noindent Thus, for $M $ large, 
\begin{equation}
\begin{array}{rcl}
\int_{m+n} \{R \ge d \varepsilon \}|\phi_d| &\le& \int_{m+n}\exp[-d^2\varepsilon^2/O(1)]\le \exp[-M^2d\log(d)/O(1) +O(d)],\\
P_d\{R \ge d \varepsilon \} &=&\int_{m+n} \{R \ge d \varepsilon \}|\phi_d|/\int_{m+n} | \phi _d | =\exp[-d/O(1)].
\end{array}
\end{equation}
\newline
\noindent {\bf Lemma 3.5}: For $M$ large enough, $\max_i P_d\{|\tilde v_i-\bar w|>\varepsilon\} =\exp[-d/O(1)]$.
\newline\newline\noindent {\bf Proof:}. 
\begin{equation}
\begin{array}{rcl}
 \mbox{ From lemma 3.5, for some } M,  P_d\{R>d\varepsilon\} &\to& 0 \mbox{ as } d \to \infty,\\
\end{array}
\end{equation}
 Define $R_{-i}=\sum_{jk,j\ne i}|\tilde v_j - \tilde w_k|^2$. Of course $R_{-i} \le R$. For $i \le m$,

\begin{equation}
R_{-i} \le d \varepsilon  \Rightarrow m\sum_k|\tilde w_k-\bar w|^2 \le d^2\varepsilon^2 
 \Rightarrow \min_k|\tilde w_k-\bar w| \le ({m \over n}+{n \over m})^{1/2} \varepsilon=\varepsilon_1.
\end{equation}
By the metric inequality, the interval $I_k=\{\tilde v| \:|\tilde v -\tilde w_k|\le \varepsilon_1\}$ on the unit circle, of length at least $2\varepsilon_1$,  is such that $|\tilde v-\bar w| \le 2\varepsilon_1$ for $\tilde v \in I$.

\noindent Letting $t_{-i}=\{t_j,j \ne i\}$, note that the conditional density of $t_i|t_{-i}$ is proportional to $\prod_k|\psi_{ik}|$. Then,
for $t_{-i}$ satisfying $R_{-i} \le \varepsilon$, and $M_2$ chosen large enough,
\begin{equation}
\begin{array}{rcl}
 \exp[-\sum_k|\tilde v_i -\tilde w_k|^2O(1)]| &\le& \prod_k|\psi_{ik}|\le \exp[-d|\tilde v_i -\bar w|^2/O(1)],\\
P_d\{|\tilde v_i-\bar w|>\varepsilon_2|t_{-i}\}&\le& \exp[-d\varepsilon_2^2/O(1)]/\int\prod_k|\psi_{ik}|dt_i  , \\
1 \ge P_d\{|\tilde v_i-\bar w| \le 2\varepsilon_1|t_{-i}\}&\ge& \exp[-d\varepsilon_1^2 O(1)] \int|\tilde v_i-\bar w| \le 2\varepsilon_1\}dt_i
/\int\prod_k|\psi_{ik}|dt_i,\\
1&\ge& 2 \varepsilon_1 \exp[-d\varepsilon_1^2 O(1)] /\int\prod_k|\psi_{ik}|dt_i\\
P_d\{|\tilde v_i-\bar w|>\varepsilon_2|t_{-i}\}&\le& \exp[-d\varepsilon_2^2/O(1)+d\varepsilon_1^2 O(1)]/2\varepsilon_1\\
&=& \exp[-d/O(1)].
\end{array}
\end{equation}
The same $M_2$ holds for all $i$ because $\mu_{jk} \approx 1$, so the $O(1)$ bounds hold for all $i$.
Finally, again with the same $O(1)$ for all $i$,
\begin{equation}
\begin{array}{rcl}
P_d\{|\tilde v_i-\bar w|>\varepsilon_2\}&=&P_d\{P_d\{|\tilde v_i-\bar w|>\varepsilon_2|t_{-i}\}\{R_{-i} \le \varepsilon\} \} 
+P_d\{P_d\{|\tilde v_i-\bar w|>\varepsilon_2|t_{-i}\}\{R_{-i} > \varepsilon\} \}\\
&\le&\exp[-d/O(1)]P_d\{R_{-i} \le \varepsilon\} +P_d\{R_{-i} > \varepsilon\} \\
\max_i P_d\{|\tilde v_i-\bar w|>\varepsilon_2\}&=& \exp[-d/O(1)].
\end{array}
\end{equation}
Now, under $P_d$, the variable $\tilde w_n$ is independent of the variable $\max_{ij}|\tilde t_i-\tilde t_j|$.
Also, if $\max_{ij}|\tilde t_i-\tilde t_j|\le \varepsilon \le 1, \tilde w_n=1$, then $\max_{i}|t_i|\le 2\varepsilon.$
(We need to constrain $\varepsilon$ so that $\max_i |t_i| \le \pi/2$ to avoid difficulties with the period $2\pi$ of the
geometric characteristic function.) Then, for some constants $M_2,M_3,M_4,M_5,M_6$,

\begin{equation}
\begin{array}{rcl}
P_d\{\max|\tilde v_i-\bar w|>\varepsilon_2\} \le \sum_i P_d\{|\tilde v_i-\bar w|>\varepsilon_2\}&=&\exp(-d/O(1))\\
P_d\{\max|\tilde w_i-\bar v|>\varepsilon_3\} &=&\exp(-d/O(1))\\
P_d\{\bar w-\bar v|>\varepsilon_4\}&=&\exp(-d/O(1))\\
P_d\{\max_{ij}|\tilde t_i-\tilde t_j|>\varepsilon_5\} &=&\exp(-d/O(1))\\
P_d\{\max_{ij}|\tilde t_i-\tilde t_j|>\varepsilon_5|\tilde w_n=1\} &=&\exp(-d/O(1))\\
P_d\{\max_{i}|t_i|\le \varepsilon_6|w_n=0\}&=&\exp(-d/O(1))
\end{array}
\end{equation}

This concludes the proof of the validity of condition V.

\newpage
\noindent {\bf 4 Equal row and column sums} \newline\newline
Consider the special case of [CM07]where the row sums are equal, and the 
column sums are equal, so that $r_i =\mu n,c_j =\mu m$. In this case 
$| V_d | =n^{m-1}n^{m-1}\sigma ^{2(m+n-1)}$ where $\sigma ^2=\mu (1+\mu )$. 
In moment calculations, it is convenient to consider the linear transform
\begin{equation}
\begin{array}{rcl}
 U&=&\sum\nolimits_j {v_j } /m+\sum\nolimits_k {w_k } /n, \\ 
 V_j &=&v_j +\sum\nolimits_k {w_k } /n,1\le j\le m, \\ 
 W_k &=&w_k +\sum\nolimits_j {v_j } /m,1\le k\le n.  
 \end{array}
\end{equation}
Note that $Q_\varepsilon (t)=1\Rightarrow | U| \le 2\varepsilon 
,| V_j | \le 2\varepsilon ,| W_k | \le 2\varepsilon .$
When $t\sim N(0,V_d^{-1} )$, the $U,V,W $ are multivariate Gaussian in $d$ dimensions 
with
\begin{equation}
\begin{array}{rcl}
 U&\sim& N(0,1/mn\sigma ^2), \\ 
 V_j &\sim& N(0,1/n\sigma ^2) \mbox{ independent },1\le j\le m, \\ 
 W_k &\sim& N(0,1/m\sigma ^2) \mbox{ independent },1\le k\le n, \\ 
 &&U,V_j -U,W_k -U \mbox{ independent }. \\ 
 \end{array}
\end{equation}

Then 
\begin{equation}
\begin{array}{rcl}
 K_d^2 &=&[-mnU^2+n\sum\nolimits_j {V_j^2 } +m\sum\nolimits_k {W_k^2 } ]\sigma^2 
, \\ 
 K_d^3 &=&[-mnU^3+n\sum\nolimits_j {V_j^3 } +m\sum\nolimits_k {W_k^3 } ]\sigma^2(1+2\mu ), \\ 
 K_d^4 &=&[-mnU^4+n\sum\nolimits_n {V_j^4 +m} \sum\nolimits_k {W_k^4 
+6\sum\nolimits_j {(V_j -U)^2} \sum\nolimits_k {(W_k -U)} ^2]} \sigma^2(1+6\sigma^2),\\ 
\end{array}
\end{equation}
\begin{equation}
\begin{array}{rcl}
E_d (K_d^3 )^2&=&3(5(m+n-1)^2-4(m-1)(n-1))(1+4\sigma ^2)/(mn\sigma ^2),\\
E_d K_d^4 &=&3(m+n-1)^2(1+6\sigma ^2)/(mn\sigma ^2),\\
\hat {P}\{S_d =0\}&=&(2\pi \sigma ^2)^{-(m+n-1)/2}m^{(1-n)/2}n^{(1-m)/2} \times\\
&&\exp ([6(m-1)(n-1)-(m^2+n^2-1)(1+1/\sigma ^2)]/12mn).
 \end{array}
\end{equation}

\noindent Dropping terms $O(1/d)$, the exponential term is  $\exp [\textstyle{1 \over 2}-(\textstyle{m \over n}+\textstyle{n \over m})(1+1/\sigma ^2)/12].$
Now the number of points satisfying $R=r,C=c$ is estimated as:
\begin{equation}
\hat {Q}(R=r,C=c)=\hat {P}\{R=r,C=c)\exp (I(P))=\hat {P}(S_d =0)[(1+\mu)^{1+\mu }\mu ^{-\mu } ]^{mn}
\end{equation}

\newpage
Using data from [CM07] , page 5,

\begin{center}
{\bf Table 1: Estimated number of contingency tables}
\end{center}

\begin{center}
{\bf with given constant row sums and constant column sums}
\end{center}

\begin{table}[htbp]
\begin{center}
\begin{tabular}{|p{73pt}|l|l|l|l|l|}
\hline
Rows & 
Cols& 
Summand mean& 
Exact& 
Edgeworth& 
[CM07]1.2 \\
\hline
10& 
10& 
2& 
1.10 10$^{59}$& 
1.12 10$^{59}$& 
1.23 10$^{59}$ \\
\hline
3& 
3& 
100/3& 
1.33 10$^{7}$& 
1.23 10$^{7}$& 
1.68 10$^{7}$ \\
\hline
3& 
49& 
49/3& 
1.01 10$^{68}$& 
4.04 10$^{147}$& 
1.25 10$^{68}$ \\
\hline
3& 
9& 
11& 
2.79 10$^{21}$& 
2.84 10$^{21}$& 
3.49 10$^{21}$ \\
\hline
18 & 
18& 
13/18& 
7.95 10$^{127}$& 
8.05 10$^{127}$& 
8.50 10$^{127}$ \\
\hline
30& 
30& 
1/10& 
2.23 10$^{59}$& 
2.23 10$^{59}$& 
2.32 10$^{59}$ \\
\hline
\end{tabular}
\label{tab1}
\end{center}
\end{table}
The hideously bad approximation at $m=3,n=49,\mbox{ mean }=49/3 $ occurs because the $n/m$ terms in the Edgeworth correction are no longer accurate. (In [CM07], Canfield and MacKay 
express their approximation as a correction to Good's joint hypergeometric 
approximation, rather than as a correction to the multivariate Gaussian 
approximation; this approach produces an estimate that does not involve $n/m$ 
terms.)

\newpage 
\noindent{\bf 5 The number of graphs with a specified degree sequence}\newline\newline
\noindent Consider a symmetric table of $0-1$ integers $X_{ij} =X_{ji} ,X_{ii} =0,1\le 
i\le n,1\le j\le n$ with given row sums $D_i =\sum\nolimits_j {X_{ij} } =d_i $. The row sums are the degrees of the undirected graph in which $X_{ij} =1$ 
corresponds to an edge between nodes $i,j$. As before we use $D_i $ for 
a random variable, $d_i $ for a particular value. The random variables $\{D_i \}$ take values on ${\{0,1,..(n-1)\}}^n$.
We wish to estimate the number of graphs with the specified degree sequence. 

The Edgeworth approximation begins with the maximum entropy distribution 
on $\{X_{ij} \}$ with expectations $ED_i =d_i $, which consists of 
independent Bernoullis with expectations $\mu _{ij} $:
\begin{equation}
 P\{X_{ij} =x\}=\mu _{ij}^x (1-\mu _{ij} )^{1-x},
\end{equation}
where 
\begin{equation}
\log (\mu _{ij} /(1-\mu _{ij} ))=\alpha _i +\alpha _j ,
\end{equation}
 and the parameters $\alpha _i $ are chosen so that 
\begin{equation}
ED_i =\sum\nolimits_j {\mu _{ij} } =r_i ,
\end{equation}
 provided that there exist $\alpha$ that solve these equations. See [BH10b] for conditions on the degree sequences for such $\alpha$'s to exist.

\noindent The conditional distribution of $\{X_{ij}\}$ 
given the degrees $\{d_i\}$ is uniform. The number of graphs with the specified 
degree sequence is
\begin{equation}
q(D)=P\{D=d\}\exp [I(P)]=P(D=d)/\prod\limits_{i<j} {(1-\mu _{ij} )}^{1-\mu _{ij} 
}\mu _{ij}^{\mu _{ij} } .
\end{equation}
The probability$P\{D=d\}$ is estimated by 
\begin{equation}
\hat P\{D =d\}=2{(2\pi )}^{-n/2}{| V_n |} ^{-1/2}\exp (-\kappa _n^3 /72+\kappa _n^4/24)
\end{equation}
determined by the first four cumulants of $D$ following the program of section 2.

The reason for the initial factor 2 is that the sum of the degrees is even; the 
lattice of all possible degree sequences has determinant $\Delta =2$. The 
characteristic function over the cube $(-\pi ,\pi ]^n$ concentrates at $t=0$ 
and also at $t=\pi $ ; the Gaussian formula for the integral near $t=0$ produces 
the same value near $t=\pi $, so the total integral is twice the formula for the integral near $t=0$. 
For nearly regular graphs, graphs whose degrees are in the ratio $1+o(n^{-1/2})$, the Edgeworth formula reproduces 
the asymptotic formula in [MW90].

Each element of $D$ is a sum of independent Bernoullis with 
expectations $\{\mu _{ij} \}$. The validity of the asymptotic estimate 
depends on the behaviour of the characteristic function of $D-d$, with 
parameters $t_j ,1\le j\le n$, setting $t_{jk}=t_j+t_k$,

\begin{equation}
\phi _n (t)=E\{\exp (it'(D-d)\}=\prod\limits_{j<k} \psi _{\mu _{jk} }  
(t_{jk} )=\prod\limits_{j<k} e^{-it_{jk}\mu _{jk} }(1+\mu_{jk} e^{it_{jk}})
\end{equation}
The cumulants $K_n^r (t)$ of $t'D$ are the 
sums of the corresponding cumulants of the Bernoullis with expectations $\mu 
_{jk} $ and parameters $t_{jk} =t_j +t_k $ ,
\begin{equation}
\begin{array}{rcl}
 K_n^2 &=&\sum\nolimits_{j<k} t_{jk}^2  \mu _{jk} (1-\mu _{jk} )=t'V_n t, \\ 
 K_n^3 &=&\sum\nolimits_{j<k} t_{jk}^3  \mu _{jk} (1-\mu _{jk} )(1-2\mu_{jk} ), \\ 
 K_n^4 &=&\sum\nolimits_{j<k} t_{jk}^4  \mu _{jk} (1-\mu _{jk} )(1-6\mu _{jk} (1-\mu _{jk} )).  
 \end{array}
\end{equation}

Then the Edgeworth approximation terms are $\kappa_n^3=E_n(K_n^3)^2, \kappa_n^4=E_n(K_n^4)$, where the expectation $E_n$ is 
under the assumption $t \sim N(0,V^{-1})$.
We show in [BH10b] that the formula (88) is valid under similar conditions for the 
contingency table case, namely that the binomial expectations are relatively 
bounded as $n$ goes to infinity.
 \newpage
\noindent {\bf 6 Regular graphs}\newline

\noindent Consider a regular graph, where the degrees all equal to $d$ .
Then $\mu =d/(n-1)$; let $v=\mu (1-\mu )$.
\begin{equation}
\begin{array}{rcl}
 V_n (i,j)&=&v(1+\delta _{ij} (n-2)), \\ 
 | V_n | &=&2(n-1)(n-2)^{n-1}v^ n ,\\ 
   V_n^{-1} (i,j)&=&{-{1 \over 2(n-1)}+\delta _{ij} \over (n-2)v}, \\ 
 E_n (t_i +t_j )(t_r +t_s )&=&{[-{2 \over n-1}+\delta _{ir} +\delta _{is} +\delta 
_{jr} +\delta _{js} ]\over (n-2)v}.  
 \end{array}
\end{equation}
These expectations may be derived directly, without inverting $V$, by noting 
that $t'V_n t\sim \chi _n^2 $ has mean $n $ and variance $2n$. The final equation is used in evaluating the third and fourth 
cumulants, using Wick's formula:
\begin{equation}
EX^4=3(EX^2)^2,EX^3Y^3=9EX^2EY^2EXY+6(EXY)^2.
\end{equation}
\begin{equation}
\begin{array}{rcl}
 \kappa _n^3 &=&\mu_n (K_n^3 )^2=6[(1-4v)^2/v][4(n-2)^2+1]/[n(n-1)], \\ 
 \kappa _n^4 &=&\mu_n K_n^4 =6(1/v-1)(n-2)/(n-1) . 
 \end{array}
\end{equation}

For n even, the estimated number of regular graphs of degree $d$ is

\begin{equation}
\begin{array}{rcl}
 \hat P\{D=d\}\exp (I(P))&=&\hat P\{D=d\}[(1-\mu )^{1-\mu }\mu ^\mu ]^{-(n-1)/2}, \mbox{ where }\\ 
 \hat P\{D=d\}&=&2{(2\pi v)}^{-n/2}{\left[2(n-1){(n-2)}^{n-1}\right]}^{-1/2} \times \\
 &&\exp \left(-{1 \over 3}\left[(1/v-4){(n-2)^2+1/4 \over n(n-1)}
 +{1 \over 4}(1/v-6){n-2 \over n-1}\right]\right),\\
\mbox{ or }\hat P\{D=d\}&=&\exp\left(-{n \over 2}\log (2\pi vn)+0.5\log 2+{5 \over 6}-{1 \over {12v}}+O({1 \over n}) \right). 
\end{array}
\end{equation}

The last formula is identical to the formula given by McKay and 
Wormald in[WM07]. The previous formula improves the accuracy for modest $n$ by 
carrying the $n-1$ and $n-2$ terms which give the exact contributions from 
the third and fourth cumulants. Note that the approximation is symmetric about the degree $d=(n-1)/2,\mu 
=1/2.$ This is as it should be, since the number of regular graphs with 
degree $d $ is the same as the number of complementary regular graphs with degree 
$n-1-d. $

\begin{center}
The estimated number of graphs is maximized at $\mu =1/2$, taking the value 
${(2^{n-2}/\pi n)}^{n/2}\exp (1/2)\sqrt 2 $.
\end{center}
This can't be too far off, since we get $2^{n(n-1)/2}$ graphs by assigning 
the $n(n-1)/2$ edges in all possible ways, and we would expect most of the 
degrees in that population of graphs to be about $d=(n-1)/2$. The other 
terms in the expression are the Gaussian correction to get the degrees 
exactly $d$, and then the Edgeworth correction that identifies 
a constant ratio departure from the Gaussian formula in the limit. 

\newpage
\begin{center}
{\bf Table 2: Log number of labelled regular graphs}
\end{center}

\begin{center}
{\bf+ error in Edgeworth approximation}
\end{center}

\begin{table}[htbp]
\begin{center}
\begin{tabular}{|p{101pt}|l|l|l|l|}
\hline
Vertices/Degree& 
3& 
4& 
5& 
6 \\
\hline
8& 
9.87+.06& 
& 
& 
 \\
\hline
9& 
& 
13.84+.04& 
& 
 \\
\hline
10& 
16.23+.10& 
18.01+.04& 
& 
 \\
\hline
11& 
& 
22.37+.05& 
& 
 \\
\hline
12& 
23.17+.14& 
26.90+.06& 
28.72+.03& 
 \\
\hline
13& 
& 
31.58+.08& 
& 
35.28+.03 \\
\hline
14& 
30.60+.18& 
36.42+.09& 
40.18+.04& 
42.04+.03 \\
\hline
15& 
& 
41.39+.10& 
& 
48.98+.03 \\
\hline
16& 
38.46+.20& 
46.49+.11& 
52.31+.06& 
56.11+.03 \\
\hline
17& 
& 
51.71+.12& 
& 
63.41* \\
\hline
18& 
46.68+.23& 
57.05+.13& 
65.04+.08& 
70.88* \\
\hline
\end{tabular}
\label{tab2}
\end{center}
\end{table}

\begin{itemize}
\item * numbers are not computed, but estimated from the Edgeworth formula
\item The approximation works best when the degree is near half the number of vertices, and gets progressively worse for fixed degree as the number of vertices increases. However, the approximations are not too bad even near the edges; for example the error for 40 vertices and degree 2 is .6 on the log scale, which is about a ratio of 2. 
\end{itemize}

 \newpage
\noindent {\bf 7 Irregular Graphs }
\newline\newline
\noindent Consider now graphs with $n_1 $ vertices of degree $d_1 $, $n_2 $ vertices of 
degree $d_2 $. The maximum entropy summands are independent Bernoullis on the edges with 
probabilities

$p_{11} $ for the edges $(i,j),1\le i<j\le n_1 $,

$p_{12} $ for the edges $(i,j),1\le i\le n_1 <j\le n_1 +n_2, $

$p_{22} $ for the edges $(i,j),n_1 <i<j\le n_1 +n_2 $.

\noindent The maximum entropy choice of the $p$'s is the unique solution , when it exists, to
\begin{equation}\begin{array}{rcl}
 (n_1 -1)p_{11} +n_2 p_{12} &=&d_1, \\ 
 (n_2 -1)p_{22} +n_1 p_{12} &=&d_2, \\ 
 \frac{p_{11} }{1-p_{11} }\frac{p_{22} }{1-p_{22} }&=&(\frac{p_{12} }{1-p_{12}})^2 . 
 \end{array}\end{equation}
The Bernoulli variances are $v_{ij} =p_{ij} (1-p_{ij} )$. 
The random degrees $D_i $ have covariance matrix $V: $
\begin{equation}\begin{array}{rcl}
 V_{ii} &=&(n_1 -1)v_{11} +n_2 v_{12} ,1\le i\le n_1 , \\ 
 V_{ii} &=&(n_2 -1)v_{22} +n_1 v_{12} ,n_1 <i\le n_1 +n_2 , \\ 
 V_{ij} &=&v_{11} ,1\le i\ne j\le n_1 , \\ 
 V_{ij} &=&v_{12} ,1\le i\le n_1 <j\le n_1 +n_2 , \\ 
 V_{ij} &=&v_{22} ,n_1 <i\ne j\le n_1 +n_2, \\ 
 | V| &=&{((n_1 -2)v_{11} +n_2 v_{12} )}^{n_1 -1}{((n_2 -2)v_{22} +n_1v_{12} )}^{n_2 -1} \times \\
&&[(2n_1 -2)v_{11} +n_2 v_{12} )((2n_2 -2)v_{22} +n_1 v_{12} )-n_1 n_2 v_{12}^2 ]. \\ 
\end{array}\end{equation}

In the case where $n_1 =n_2 =n/2,d_2 =n-d_1 -1,n/4<d_1<3n/4,$ 
then $p_{12} =1/2,p_{11} =1-p_{22} =(d_1 -{1 \over 
4}n)/({1 \over 2}n-1),v_{11} =v_{22} ,v_{12} ={1 \over 
4},$ and the covariances of the $t_{ij} =t_i +t_j $ needed for $\kappa _n^3 
,\kappa _n^4 $ are:
 
\begin{equation}\begin{array}{rcl}
A&=&({1 \over 2}n-2)v_{11} +n/8 \\ 
 Q&=&((n-2)v_{11} +n/8)^2-(n/8)^2, \\ 
 V_{ii}^{-1} &=&1/A+V_{12}^{-1} \\ 
 V_{ij}^{-1} &=&\left\{n/16-v_{11} \left[(n-2)v_{11} +n/8\right]\right\}/(AQ),1<i <j\le n/2, \\ 
 V_{ij}^{-1} &=&-{1 \over 4}/Q,1\le i\le n/2,n/2<j\le n, \\ 
  | V| &=&(({1 \over 2}n-2)v_{11} +n/8)^{n-2}Q ,\\ 
 N_{ij} &=&\{1\le i\le n/2\}\{n/2<j\le n\}+\{n/2<i\le n\}\{1\le j\le n/2\}, \\ 
 E_nt_{ij} t_{kl} &=&4V_{12}^{-1} +(\delta _{ik} +\delta _{il} +\delta _{jk} 
+\delta _{jl} )/A+4(V_{1n}^{-1} -V_{12}^{-1} )\{N_{ik} +N_{il}+N_{jk} +N_{jl} \} \\ 
 \end{array}\end{equation}

\begin{equation}\begin{array}{rcl}
 K_n^3 &=&v_{11} (1-2p_{11} (\sum\nolimits_{1\le j<k\le n/2} {t_{jk}^3 } 
-\sum\nolimits_{n/2<j<k\le n} {t_{jk}^3 } ) \\ 
 K_n^4 &=&v_{11} (1-6v_{11} )(\sum\nolimits_{1\le j<k\le n/2} {t_{jk}^4 } 
+\sum\nolimits_{n/2<j<k\le n} {t_{jk}^4 } )-{3 \over 
8}\sum\nolimits_{1\le j\le n/2<k\le n} {t_{jk}^4 } \\ 
 \end{array}\end{equation}
The Gaussian approximation: 
\begin{equation}
\hat Q_G\{D=d\}=2(p_{11} \log p_{11} +p_{22} \log p_{22} )^{-n(n-2)/4}(\log 2)^{-n^2/4}(2\pi )^{-n/2}| V| ^{-1/2}.
\end{equation}
The initial 2 is the determinant of the lattice of possible degree 
sequences. The second term is the contribution from the Bernoulli 
probabilities, the exponential value of the entropy. The last term is the 
Gaussian contribution for the probability that $D=d$. The Edgeworth correction multiplies by the factor\newline
$\exp (-\kappa _n^3 /72+\kappa _n^4 /24)$ computed by
$\kappa _n^3 =E_n(K_n^3 )^2,\kappa _n^4 =E_n K_n^4 $ where the expectation is taken under the assumption $t \sim N(0,V^{-1})$.

\newpage
\begin{center}
{\bf Table 3: Log number of graphs with irregular degree sequences}
\end{center}

\begin{table}[htbp]
\begin{center}
\begin{tabular}{|p{110pt}|l|l|l|}
\hline
Degree Sequence& 
Exact& 
Gauss& 
Edgeworth \\
\hline
44443333& 
9.59& 
10.22& 
9.64 \\
\hline
666666555555& 
28.45& 
29.03& 
28.46 \\
\hline
77777774444444& 
24.21& 
24.83& 
24.33 \\
\hline
\end{tabular}
\label{tab3}
\end{center}
\end{table}

\noindent The Edgeworth formula is significantly more accurate than the Gaussian 
formula. The Edgeworth formula is more accurate when the degrees are nearly 
equal.


\newpage 
\begin{thebibliography}{8}

\bibitem[B09]{B09} A. Barvinok, \emph{ Asymptotic estimates for the number of contingency tables, 
integer flows,and volumes of transportation polytopes}, Int. Math. Res. Notices \textbf{2009 } 
(2009), 348--385.

\bibitem[BH09]{BH09} A. Barvinok and J.A. Hartigan, \emph{An asymptotic formula for the number of non-negative integer matrices with prescribed row and column sums}, arXiv:0910.2477 ,2009
   
\bibitem[BH10a]{BH10a} A.Barvinok and J.A. Hartigan, \emph{Maximum entropy Gaussian approximations for the number
of integer points and volumes of polytopes}, Advances in Applied Mathematics \textbf{45} (2010), 252–-289

\bibitem[BH10b]{BH10b} A.Barvinok and J.A. Hartigan, \emph{The number of graphs and a random graphs with
a given degree sequence}, arXiv:1003.0356 ,2010

\bibitem[CM05]{CM05} E.R.Canfield and  B.D.McKay, \emph{Asymptotic enumeration of dense 0-1 
matrices with equal row sums and equal column sums}, Electronic J. Combin., \textbf{12} (2005), 29. 

\bibitem[CM07]{CM07} E.R.Canfield and  B.D.McKay, \emph{Asymptotic enumeration of Contingency 
Tables with Constant Margins } arXiv:0703.600v1 (2007).

\bibitem[CGM08]{CGM08} E.R. Canfield, C.Greenhill, and B.D.McKay, \emph{Asymptotic enumeration 
of dense 0-1 matrices with specified line sums}, J. Combin. Th., Ser. A \textbf{115} (2008), 32--66. 

\bibitem[DE85]{DE85}  P.Diaconis and B.Efron, \emph{Testing for independence in a two-way 
table: new interpretations of the chi-square statistic. With discussions and 
with a reply by the authors}, Ann. Statist. \textbf{13} (1985), 845--913.

\bibitem[D87]{D87}  J.Duoandikoetxea, \emph{Reverse H\"{o}lder Inequalities for Spherical 
Harmonics}, Proc. Am. Math. Soc. \textbf{101} (1987), 487--491.

\bibitem[G76]{G76} I.J.Good, \emph{On the application of symmetric Dirichlet distributions and 
their mixtures to contingency tables},Ann.Statist. \textbf{4} (1976), 1159--1189.

\bibitem[GMW06]{GMW06} C.Greenhill, B.D.McKay, and X.Wang, \emph{Asymptotic enumeration of 
sparse 0-1 matrices with irregular row and column sums}, J. Combinatorial 
Theory, Ser. A, \textbf{113} (2006), 291--324. 

\bibitem[J57]{J57} E.T.Jaynes, \emph{Information Theory and Statistical Mechanics, Physical 
Review} \textbf{106} (1957),620--630.


\bibitem[KT03]{KT03} S.J.Kathman and G.R.Terrell, \emph{Poisson approximation by constrained exponential tilting},
 Statistics \& Probability Letters  \textbf{61} 2003, 83--89

\bibitem[K06]{K06} J.E.Kolassa ,\emph{Series Approximation Methods in Statistics, 3rd Edition }(Lecture Notes in Statistics 88), Springer, New York, 2006.

\bibitem[MW90]{MW90} B.D.McKay, and N.C.Wormald, \emph{Asymptotic enumeration by degree 
sequence of graphs of high degree}, European J. Combin. \textbf{11} (1990), 565--580

\bibitem[MG]{MG} B.D.McKay, and C.S.Greenhill, \emph{Asymptotic enumeration of sparse 
nonnegative integer matrices with specified row and column sums},  
Adv. in Appl. Math. \textbf{41} (2008), 459--481.

\end{thebibliography}
\end{document}